\documentclass[a4paper,12pt]{spieman}  
\usepackage{amsmath,amsfonts,amssymb}
\usepackage{graphicx}
\usepackage{setspace}
\usepackage{tocloft}
\usepackage{enumerate}
\usepackage[shortlabels]{enumitem}
\usepackage{multicol}
\usepackage{multirow}
\usepackage{booktabs}
\usepackage{xcolor}
\usepackage{soul}

\title{Simulating the optical alignment of the Multi-Conjugate Adaptive Optics module for the Extremely Large Telescope}

\author[a,*]{Mauro Patti}
\author[a]{Matteo Lombini}
\author[b]{Edoardo Maria Alberto Redaelli}
\author[a]{Emiliano Diolaiti}
\affil[a]{INAF $-$ Osservatorio di Astrofisica e Scienza dello Spazio di Bologna (Italy)}
\affil[b]{INAF $-$ Osservatorio Astronomico di Brera, Merate (Italy)}

\cftpagenumbersoff{figure}
\cftpagenumbersoff{table} 
\begin{document} 
\maketitle
\begin{abstract}
Adaptive optics instruments for the future extremely large telescopes are characterised by advanced optical systems with diffraction-limited optical quality. Low geometric distortion is also crucial for high accuracy astrometric applications. Optical alignment of such systems is a crucial step of the instrument integration. Due to relative inaccessibility of these giant instruments, automatic alignment methods are also favoured, to improve the instrument availability after major events, such as extraordinary maintenance. The present paper describes the proposed alignment concept for these systems: the notable example which is analysed here is the case of the Multi-conjugate Adaptive Optics RelaY (MAORY) for the future Extremely Large Telescope.
The results of ray-tracing simulations carried out to validate the method are discussed in detail, covering the error sources which could degrade the alignment performance.  
\end{abstract}


{\noindent \footnotesize\textbf{*}Mauro Patti,  \linkable{mauro.patti@inaf.it} }

\begin{spacing}{1}   

\section{Introduction}
\label{intro}
The future generation of extremely large telescopes~\cite{gilmozzi2007european, sanders2013thirty, johns2008giant} will host instruments which need large optics if a large Field of View (FoV) shall be covered. The Adaptive Optics~\cite{babcock1953possibility} (AO) architecture is essential to exploit the large telescope aperture and achieve various science goals which rely on angular resolution. AO instruments are complicated and need a fair number of optical elements to re-image the telescope focal plane with diffraction-limited optical quality. The complexity of these optical system is due to many aspects such as the off-axis, not rotationally symmetric shape of optical elements and a high number of degrees of freedom (DoFs). Large, complex and very demanding in terms of performances, the new generation of instruments will require time consuming and difficult alignment procedures. In many cases, it is difficult to divide the system into parts that can be aligned separately, testing the performance of the resulting subsystem at every step of the integration process by means of several optical tools. Due to relative inaccessibility of these instruments, which will be installed at the focal stations of the telescopes, maintenance is also an issue: automatic alignment procedures should be favoured, in order to minimise the instrument down-time after major maintenance events, e.g. after optics re-coating or possible slight misalignment after an earthquake. The issues discussed above motivate a different approach to optical alignment which is based on automatic methods and numerical algorithms~\cite{rimmer1990computer}. This paper proposes a reverse optimization algorithm which uses the merit function minimization of commercial software for optical systems design. The development of such kind of algorithms has been originally treated by M. Egdall~\cite{egdall1985manufacture} as a possible solution for the alignment of large aspheric mirrors.\\ Optical systems, including components which are off-axis sections of rotationally symmetric parent surfaces, are not trivial to align. For example, the alignment of a two off-axis mirror system by minimizing the root mean square (RMS) wavefront error (WFE) at the centre of the FoV, could exhibit large aberrations at the edge of the field~\cite{wilson1996aberration}. In the last three decades, many authors have applied reverse optimization algorithms to study the alignment of non-rotational symmetric optical systems~\cite{jeong1987auto, jeong1989simultaneous,kim2005reverse}. One of the first successful applications, by using reverse optimization and a Hartmann screen, has been achieved by A. Lundgren and M. Wolfe~\cite{lundgren1991alignment}. However, they stated that even if an acceptable alignment was achieved, the system did not perform as well as predicted. Possible explanations for this mismatch were errors that made the real system deviate from the nominal one (e.g. errors in the actual shape of the mirrors). The simulation, described in this paper, considers also the alignment and manufacturing errors in order to predict the performance during the real system integration phase.\\
The design of future Multi-Conjugate Adaptive Optics (MCAO) instruments~\cite{diolaiti2016maory,crane2018nfiraos} for extremely large telescopes aims to enhance the astrometric performance of current MCAO-assisted cameras~\cite{patti2019gems}. For this reason, the alignment of such instruments has to provide not only an excellent optical quality but also a very low geometric distortion. As a test case, the reverse optimization technique is applied to MAORY~\cite{diolaiti2016maory}, the MCAO relay of the Extremely Large Telescope (ELT)~\cite{gilmozzi2007european}, exploring (as it has never been done before) a wide range of error sources and demonstrating the reliability of the algorithm to bring the optical system close to its nominal performance in terms of RMS WFE and geometric distortion.\\ The paper is organized as follows: in Section~\ref{sec:2}, the alignment concept and the logic behind the method are described.
In Section~\ref{sec:3}, a brief introduction to MAORY is given along with its nominal performances in terms of optical quality and geometric distortions. The implementation of the alignment method to MAORY is described in Section~\ref{sec:4}. The verification of the alignment method has been done by simulation and it is presented in Section~\ref{sec:simulation}. A Monte Carlo approach has been used to simulate a range of optics misalignments considering also different error sources that affect the measurements. These are described in detail in Section~\ref{sec:errors} where each possible error source is analyzed separately and the complete results are shown at the end.
\section{Optical alignment concept}
\label{sec:2}
The sensitivity analysis is the first step in optical design to identify components which are highly sensitive to certain errors, such as tilts or decenters. The analysis is also valuable for finding the optimum number of compensators and the required range of adjustment which are necessary to reduce the performance degradation of an overall misaligned optical system.  
The compensators should work in combination with a metrology system that provides the necessary information to change the optical path according to estimated misalignments. A possible method is based on Singular Value Decomposition (SVD) and regularization of the sensitivity matrix.\\ Wavefront data can be expressed in polynomial form and Zernike polynomials~\cite{noll1976zernike} are commonly used to describe wavefront aberrations over a circular aperture. Since the future generation of large instruments will have approximately circular apertures, we will refer to Zernike coefficients for the rest of the paper.  
The DoFs sensitivities are measured in terms of Zernike coefficients across the field. By perturbing each of the individual DoFs for each optical surface, the change of each Zernike coefficient is recorded in a matrix whose rows and columns are respectively the total number of Zernike values per field point and the total number of DoFs. This sensitivity matrix includes all the necessary information to simulate the alignment using the best set of compensators. It is defined as:
\begin{equation}
S_{ij} = \frac{\partial Z_{i}}{\partial x_{j}}
\end{equation}
where the WFE is decomposed into $n$ Zernike polynomials ($Z_{i}$ ; $0<i<n$) and the root sum squared (RSS) of $Z_{i}$ is the total RMS WFE. $\partial Z_{i}$ is measured from the nominal configuration as $(Z_{perturbed,i} - Z_{nominal,i})$ while $\partial x_{j}$ is the $j^{th}$ DoF perturbation. Once the coupling between DoFs and their effect on Zernike coefficients is understood, a solution on misalignments could be found.
During the alignment phase, the Zernike coefficients have to be measured for different field points by means of one or more wavefront sensors (WFSs) or they could be estimated by imaging techniques such as the phase diversity~\cite{gonsalves1982phase, mugnier2006phase}. Following an earthquake or any other intervention, as well as during the alignment phase, an instrument needs a set of reference sources at the so-called object space and one or more detectors at the image space to perform wavefront measurements by phase diversity. The AO instruments can use the WFS which requires a single image to derive the wavefront, unlike the phase diversity that needs at least two images (in-focus and/or intra/extra focal images).\\   
Let us consider the Zernike coefficients as a metric to evaluate the system performances ($MF$) and all the available DoFs listed in a vector ($x_{c}$). The relation between the two can be written:
\begin{equation}
MF = f (x_{c})
\end{equation}
The goal is to find the best $x_{c}$ that minimize $MF$. This relation could be non-linear and interdependencies between DoFs could occur. The method followed to solve the problem is based on the one described by Chapman et al. ~\cite{chapman1998rigorous} where, through the SVD of the sensitivity matrix, the eigenvectors define the best set of DoFs used to compensate for misalignments.
The sensitivity matrix is decomposed as follows:
\begin{equation}
S=UMV^T
\end{equation}
where $U$ and $V$ are column orthogonal matrices and $M$ is a diagonal matrix containing the singular values listed in decreasing order along the diagonal. They are always positive numbers and define the strength of the type of aberrations that result from changes in the DoFs. The most sensitive aberration due to a set of DoFs misalignment has the largest singular value. Columns of $U$ are Zernike coefficients introduced by changes in DoFs listed in columns of $V$, the strength of which is given by the M singular values. The goal is to find the most linearly independent set of compensators within the sensitive subspace of aberrations. This is done by truncating the sensitivity matrix until the sum of residuals of the new singular values are below a given threshold. Looking at the singular values of $M$ in decreasing order, there is a well-defined gap at some point. The threshold has been chosen to remove all the singular values below the gap. 
The truncated sensitivity matrix $S_{tr}$ contains less singular values close to zero and it is well conditioned. The converging solutions in terms of changes in DoFs that minimize the WFE is the following:
\begin{equation}
\partial x = -V_{tr}\: M_{tr}^{-1}\:  U_{tr}^{T} \: Z
\end{equation}
where $(-V_{tr}\: M_{tr}^{-1}\:  U_{tr}^{T})$ is the pseudo-inverse SVD of $S_{tr}$ and $Z$ is the matrix of Zernike aberrations. 
This method gives accurate solutions only when the misalignments of optical components are located within the linear regime of sensitivities (i.e. perturbations around the nominal positions). In cases of large misalignments, a non-linear algorithm for minimising MF must be used. 
The method implemented in this paper is based on the damped least squares algorithm of Zemax\textsuperscript{\textregistered} ray-tracing software and it has been named Zemax Reverse Optimization (ZeRO)~\cite{patti2018precise}. In general, the method can be used with any ray-tracing software that implements a programming tool. Once a merit function for system performances is defined, the goal is to minimize the function:
\begin{equation}
\label{eq:5} 
MF^2= \frac{\displaystyle\sum_{i=1}^{n}W_{i }(V_{i}-T_{i})^2 }{\displaystyle\sum_{i=1}^{n}W_{i}}
\end{equation}
where $V_{i}$ is the current value of the $i^{th}$ Zernike coefficient, $T_{i}$ the $i^{th}$ target value, and $W_{i}$ the weight. The best set of compensators (i.e. variable parameters of the MF) is defined by the method previously described, and the perturbations required to correctly align the system, are found by minimizing the $MF$.\\  By using ZeRO, if we want to align both the focal plane and the exit pupil to their nominal positions, the measurements on Zernike coefficients are not enough. Additional observables, that could be used as targets of the $MF$, are the images centroids and the chief ray angles (e.g. by using a Hartmann screen~\cite{jeong1989simultaneous}). If a Shack-Hartmann WFS (SH-WFS) is available, the flux within the sub-apertures, instead of the chief ray angle, can be used as observable for the exit pupil position.
\begin{figure}[!b]
\begin{center}
\begin{tabular}{c}
  \includegraphics[height=5.5cm]{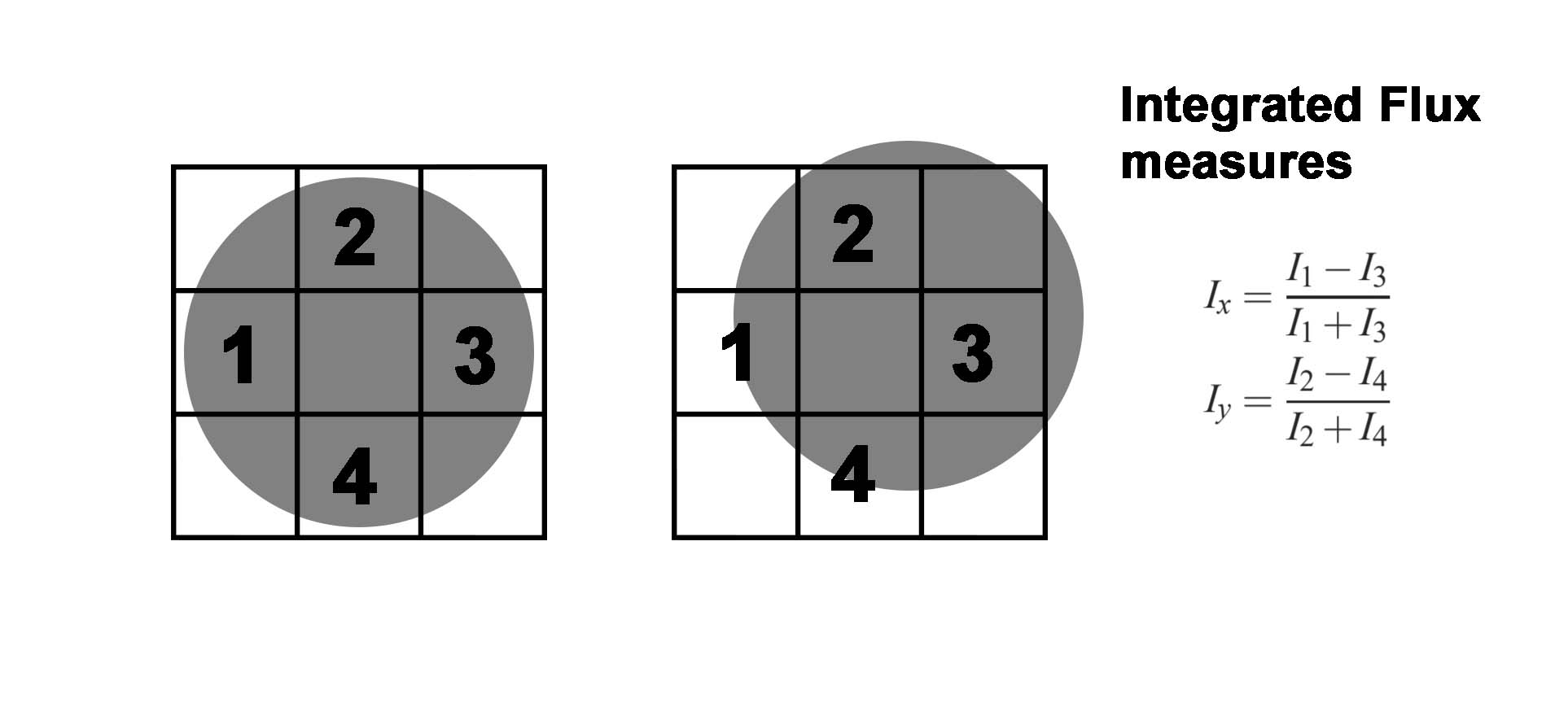}
    \end{tabular}
\end{center}
\caption{Schematic representation of the pupil shift measurement using the intagrated flux of the SH sub-apertures located at the edge of the lenslet array.}
\label{fig:5}       
\end{figure}
\\The flux within the Shack-Hartmann (SH) sub-apertures is a direct measure of the $x,\: y$ coordinates of the pupil~\cite{montagnier2007pupil} as shown in Figure~\ref{fig:5}.
Considering four areas at the edges of the $x,\: y$ axes of the lenslet array, the integrated flux of each area is defined as $I_m$ ($m = 1, ..., 4$). If $d$ is the length of the sub-aperture side which defines the area, the $x,\: y$ coordinates shift of the pupil are:
\begin{equation}
x=\frac{(I_1-I_3)}{(I_1+I_3 )} d\quad;\quad y=\frac{(I_2-I_4)}{(I_2+I_4 )} d    
\end{equation}                                                
Thus, if we have an array of sources at the object space, the $V_i$ and $T_i$ values of equation~\ref{eq:5} are the Zernike coefficients of their wavefronts, the $x,y$ coordinates of their exit pupils and the images centroids coordinates at the focal plane. In principle, the observables of one source at the FoV centre are enough to align the exit pupil and focal plane, but more field points may be used to better constrain the merit function and reduce the errors.\\
The ZeRO alignment algorithm is based on a reference optical design which is the model of the ray-tracing software. This model is a perfectly aligned system and the observables, measured from the real misaligned system, are set as optimization targets. The chosen compensators, as described above, are set as variables of the optimization routine. The ZeRO outputs are the compensators motions that minimize the $MF$ and that should be applied in a reverse way to achieve the alignment.\\The future generation of instruments will use a laser tracker (LT) as main tool of the integration phase. The LT will be used as the global metrology system to survey and align every interface and to assemble and integrate every subsystem to the instrument main structure. The LT is a portable coordinate measuring machine that permits linking the positions of the optical elements at the instrument main structure. Spherically mounted retro-reflectors (SMRs) are the reference points used by the LT. The accuracy of the LT decreases with the distance between the LT and the measured objects and with the angle that the LT head needs to assume in order to 'see' all the SMRs on the opto-mechanics~\cite{burge2007use}. As we will show at the end of the paper, the MAORY integration by only using a LT is not enough to achieve the requested performance. Thus, the ZeRO routine has to be considered.
\section{MAORY}
\label{sec:3}
MCAO is a technique first proposed by Beckers~\cite{beckers1993adaptive} to compensate the atmospheric turbulence effects over a wide field of view (FoV), by introducing more correcting elements (Deformable Mirrors, DMs) along the optical path, optically conjugated to different altitudes from the telescope pupil.
MAORY will be placed on the straight-through focus of the ELT Nasmyth platform. The MCAO wavefront sensing scheme is based on six Laser Guide Stars (LGSs) and three Natural Guide Stars (NGSs). In particular, high-order wavefront sensing is performed by using LGSs while low-order wavefront sensing is performed by using NGSs to overcome the unreliability of the LGSs measurements for these modes~\cite{ellerbroek2003mcao}. 
\subsection{Opto-mechanical design}
\begin{figure}[!h]
\begin{center}
\begin{tabular}{c}
  \includegraphics[height=5.5cm]{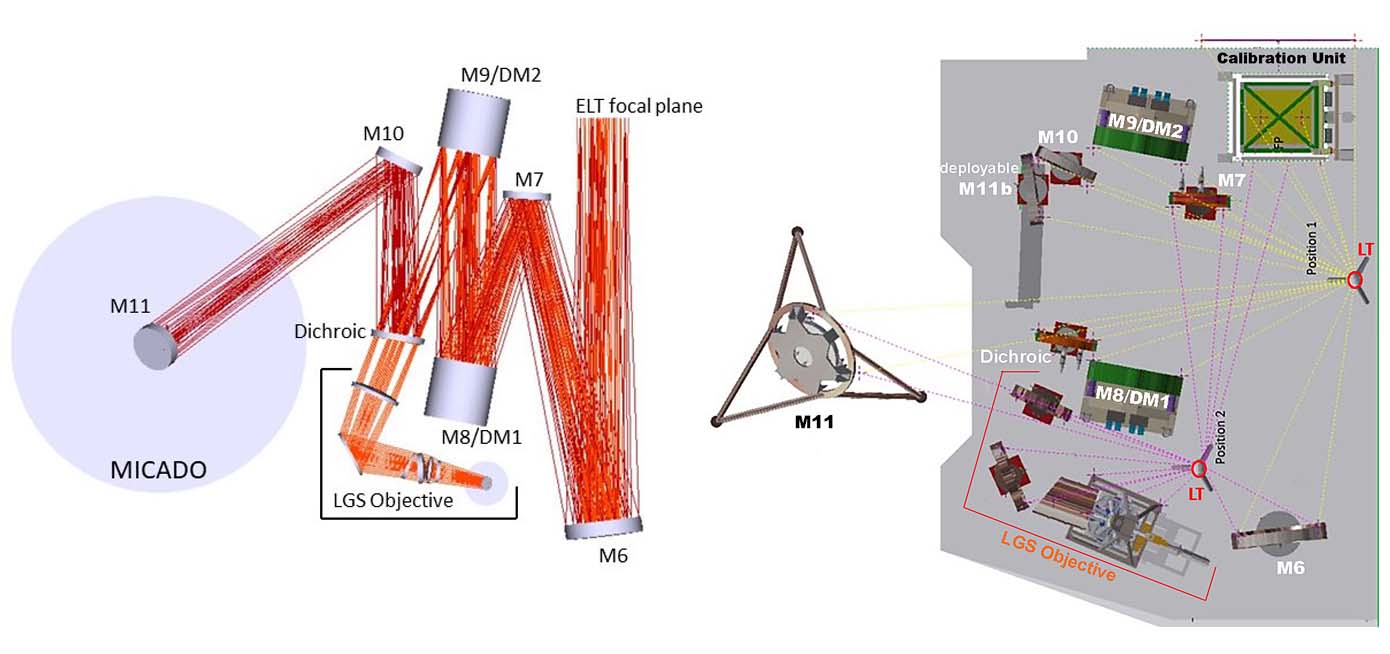}
  \\
  (a) \hspace{5.1cm} (b)
  \end{tabular}
\end{center}
\caption{(a) PFR layout and (b) realistic view of the MAORY bench with optomechanics. M11 is mechanically connected to MICADO (not shown). Two positions for the laser tracker (LT) measurements are shown.}
\label{fig:1}       
\end{figure}

The Post-Focal Relay Optics (PFR) sub-system of MAORY~\cite{lombini2018optical} re-images the telescope focal plane to the exit ports. The sub-system contains the following channels:
\begin{itemize}
\item	Main path optics (MPO), which relay the telescope focal plane to the exit ports for the science instruments: the Multi-AO Imaging Camera for Deep Observations (MICADO)~\cite{davies2010micado} and a $2^{nd}$ instrument as yet undefined;
\item	LGS objective, which creates an image plane for the LGSs, used by the LGS wavefront sensor (WFS) sub-system to measure in real-time the high-order wavefront aberrations for the MCAO mode of MAORY.
\end{itemize}
Light separation between the two channels is accomplished by a LGS dichroic beam-splitter, which splits the light reflecting the science light path and transmitting the LGS light path.
The baseline optical design of the MPO consists of 6 mirrors plus the dichroic beam-splitter. The mirrors are usually labelled from M6 (the first mirror in MAORY, after M5 in the telescope) to M11. All the mirrors with optical power, except the DMs, are off-axis section from a larger parent surface.
In detail, the optical design of the MPO, shown in Figure~\ref{fig:1}, consists of:
\begin{itemize}
\item	Concave M6 and convex M7, which produce a pupil image of the appropriate size (a concave mirror alone would not be enough, given the available space between the telescope focal plane and the edge of the Nasmyth platform); 
\item	Two concave DMs (M8 and M9), with the same optical power, optically conjugated to the required ranges from the telescope entrance pupil. In a partial implementation of the instrument, the deformable mirror M9 would be replaced by a rigid (i.e. non-adaptive) mirror. For the scope of this paper, they should be intended as rigid mirrors;
\item	LGS dichroic, close to the pupil image; in the baseline design, this component is assumed to reflect the science light;
\item	Convex M10, the last mirror with optical power before the exit focal plane;
\item	Flat $45^\circ$-tilted mirror (M11), which folds the light to the gravity-invariant port for MICADO.
\end{itemize}
The second instrument port is achieved by inserting a flat folding mirror (M11b) between M10 and M11, and ensures the delivery of the nominal optical quality focal plane as for the MICADO port.\\
The focal ratio at the MICADO focal plane is 17.745, basically equal to the ELT focal plane one. The exit pupil delivered to MICADO is located at about 9120 $mm$ before the MICADO focal plane. A summary of the MPO specifications is listed in Table~\ref{tab:1}.
\begin{table}[!h]
\caption{MPO general specifications. Geometric distortion is defined in Section~\ref{sec:2}}
\label{tab:1}       
\begin{center}   
\begin{tabular}{ll}
\toprule
\multicolumn{1}{c}{\textbf{Item}} & \multicolumn{1}{c}{\textbf{Value}} \\ \midrule
NGS patrol FoV (also called technical FoV) & 180 arcsec diameter \\
MICADO science FoV & Up to 75 arcsec diameter \\
Transmitted wavelength range at exit focal plane & 0.6 $\mu m <\lambda<$ 2.5 $\mu m$ \\
Exit focal plane curvature & Flat \\
Exit focal ratio & F/17.745  \\
Exit pupil distance & 9120 $mm$ (towards telescope)  \\
Exit pupil diameter & 510 $mm$ \\
Exit pupil blur over the MICADO FoV & 0.54$\%$ of its diameter  \\
Exit pupil eccentricity over the MICADO FoV & 0.046$\%$ of its diameter  \\
RMS WFE over the MICADO FoV & $<$ 40 $nm$ \\
Geometric distortion over the MICADO FoV & $<$ 0.3 $mas$
\end{tabular}
\end{center}
\end{table}
\\In general, each optical element within the MPO consists of an optical component (reflective or refractive) and its cell: the optical components are assumed here to be delivered already integrated into their cells. The assembly (optical component + cell) is supported by an optical mount.\\ Every optical mount includes the actuators for the adjustment of the optical element. Some of them are motorized and will be used by ZeRO during the alignment, the others are manual. In Section~\ref{sec:4}, the details for choosing motorised compensators is described. The motion of these compensators are provided with two motors which permit the rotations in azimuth and elevation by means of two kinds of gimbals.
\begin{figure}[!ht]
\begin{center}
\begin{tabular}{c}
  \includegraphics[height=5.5cm]{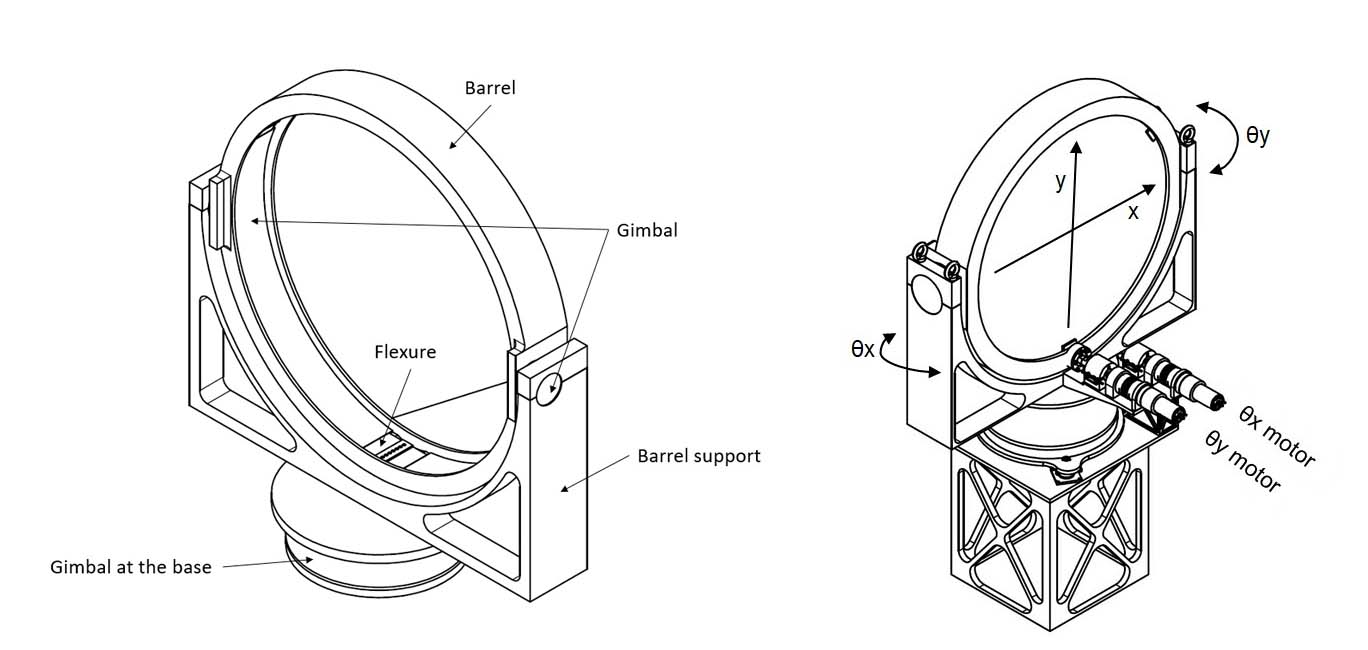}
    \\
  (a) \hspace{5.1cm} (b)
    \end{tabular}
\end{center}
\caption{Mountings view (a) with motors and structure (b). Two motors permit the elevation and azimuthal rotation of the the optomechanics with respect to its local reference origin.}
\label{fig:2}       
\end{figure}
This layout, shown in Figure~\ref{fig:2}, presents the advantage that the pivot point corresponds with the center of the optical element (surface vertex). The motors have two different positions on the mount: 
\begin{enumerate}
\item Attached to the cell to provide the elevation ($\theta_{y}$). 
\item Attached to the gimbal at the base to provide the azimuth ($\theta_{x}$). \\
\end{enumerate}
\subsection{Optical performance}
\begin{figure} [!h]
\begin{center}
\begin{tabular}{c}
  \includegraphics[height=5.5cm]{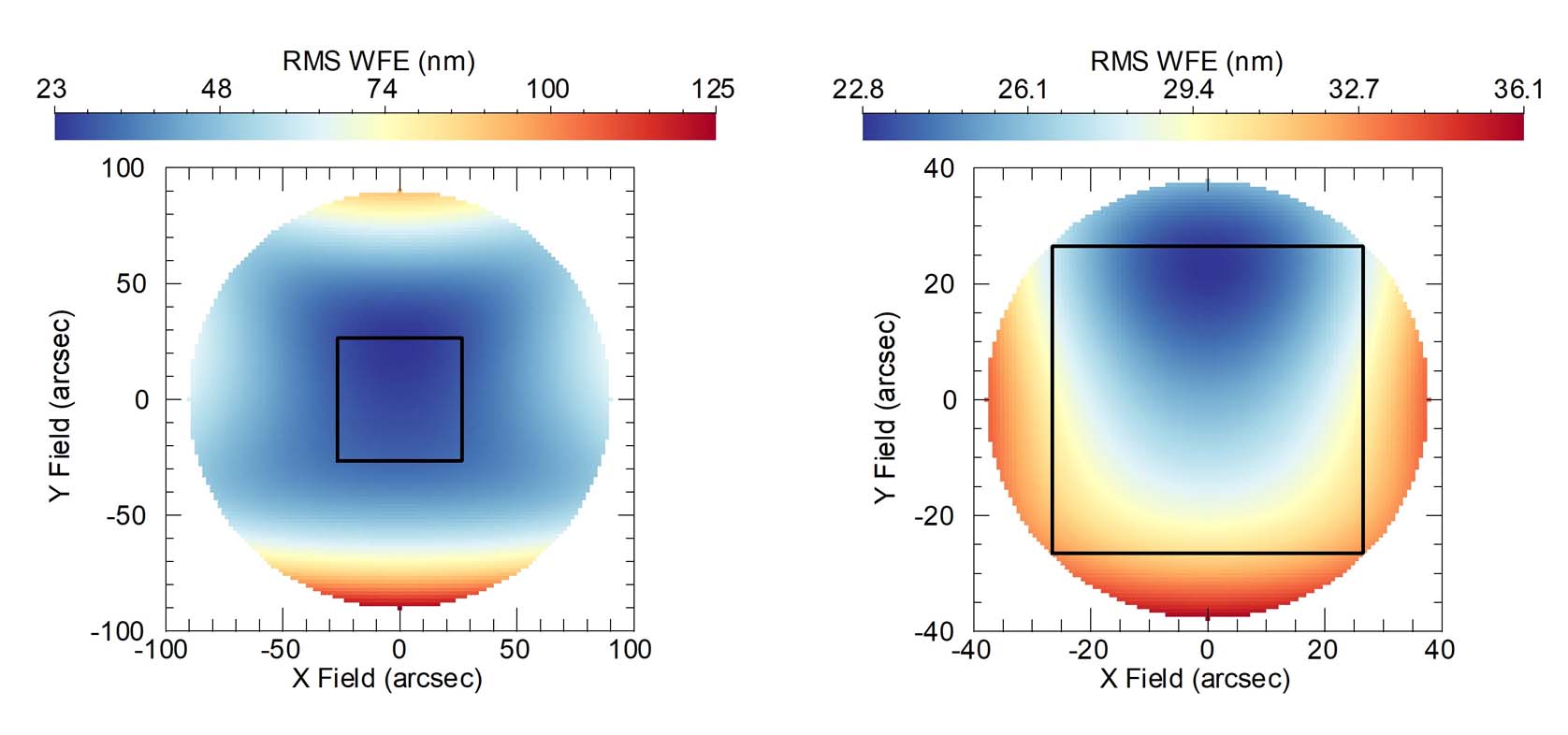}
    \\
  (a) \hspace{5.1cm} (b)
    \end{tabular}
\end{center}
\caption{(a): WFE (in nanometers) at the exit port of MAORY for the nominal optical design (i.e. without degradation from manufacturing and alignment errors; nominal telescope design included). The NGS patrol FoV is enclosed between the MICADO FoV of 53''x53'' (black square) and a 180'' diameter circle. (b): Star centroids movement for a single astrometric image in the circle containing the MICADO FoV (black square). The longest integration time for a single astrometric image is 120 seconds, corresponding to 2.6$^\circ$ field rotation at maximum.}
\label{fig:3}       
\end{figure}

\begin{figure}[!t]
\begin{center}
\begin{tabular}{c}
  \includegraphics[height=5.5cm]{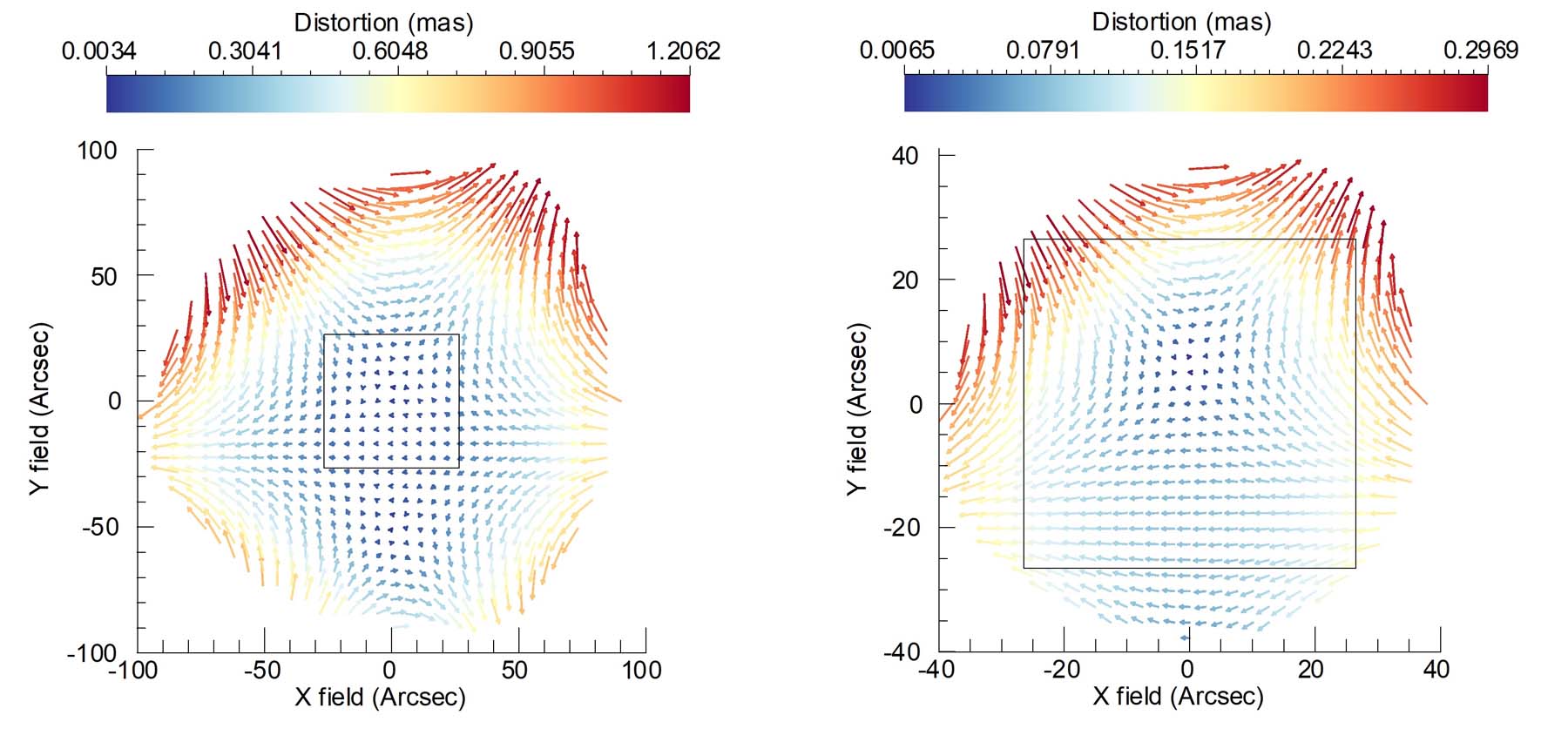}
    \\
  (a) \hspace{5.1cm} (b)
    \end{tabular}
\end{center}
\caption{Distortion map at the exit port within the maximum integration time for narrow band astrometric observations (without degradation from manufacturing and alignment errors; nominal telescope design included). (a): Distortion map in the NGS patrol FoV. (b): Distortion map in the circle containing the MICADO FoV (black square).}
\label{fig:4}       
\end{figure}

MAORY requirements are driven by the science cases which request high angular resolution and astrometric accuracy. System requirements have been broken down into sub-systems requirements which lead the sub-systems design. Strictly speaking about optical performance, the design of the MPO shall satisfy the following requirements:
\begin{itemize}
    \item Diffraction limited performance in J band over the full scientific FoV.
    \item Low geometric distortion to constrain the Point Spread Function (PSF) blur within 1/10 of its full width at half maximum (FWHM) during a single-exposure image. The details are described below.
\end{itemize}
The nominal RMS WFE at the exit focal plane of the MPO is shown in Figure~\ref{fig:3}. In particular, the WFE is below 40 $nm$ RMS over the MICADO FoV. In order to satisfy the MPO requirements, the error budget allows an additional 40 $nm$ of RMS WFE after the manufacturing, assembly and integration of the instrument.
\\MAORY/MICADO astrometric observations are taken at different epochs in order to detect motions of astronomical sources.
The current MAORY optical design has distortions of up to tens of milli-arcsec at the edges of the field. Geometric distortions from the MPO are asymmetric. As the imaged sky rotates with respect to the optics during an exposure, the PSF of a point-like source on the exit focal plane follows an imperfect trajectory, which may be described as an arc of a circle, with small perturbations due to the geometric distortions from the optics. Field de-rotation only compensates for the circular part of the trajectory. On the MICADO FoV, the residual motion of the PSF, when integrated over a finite exposure time, translates into PSF blur.
The distortion analysis presented in this paper considers the maximum integration time for narrow band astrometric observations at 80$^\circ$ of telescope elevation, i.e. 10$^\circ$ from zenith.
At the time of writing, the requirement on geometric distortions from the optics is described in the following~\cite{patti2018maory}: 
\begin{itemize}
\item	T $\approx$ 120 sec. It is the maximum integration time for narrow band astrometric observations; 
\item	A $\approx$ 13.7 x (1/cos(80$^\circ$)) $\approx$ 79 arcsec/s. It is the maximum de-rotator angular velocity;
\item	T x A $\approx$ 2.6$^\circ$. It is the maximum field rotation within a single exposure image.
\end{itemize}
During this rotation, the FWHM of the long-exposure PSF due to the MAORY optics shall not increase by more than 1/10 of its nominal value.
For wavelength $\lambda=$1 $\mu m$, the amount of distortion shall be $<$ 2.4 $mas$ at the MAORY exit focal plane on the MICADO FoV (F/17.74 focal ratio).
The positions of a set of $N$ test stars placed over a regular grid have been used to evaluate the geometric distortion. Let us define $X_{1n}  ,Y_{1n}$, ($n = 1, ..., N$), the initial coordinates of star centroids and $X_{2n}, Y_{2n}$, the coordinates of the star centroids after the maximum astrometric exposure time (i.e. 2.6$^\circ$ FoV rotation and counter-rotation angle). The PSF centroids move, because of geometric distortions, by $\Delta X_n=(X_{2n}-X_{1n})$ and $\Delta Y_n=(Y_{2n}-Y_{1n})$. The vector sum of $\Delta X_n$ and $\Delta Y_n$ is the geometric distortion experienced by the $n^{th}$ test star and shown in Figure~\ref{fig:4}. Over the full scientific FoV, the maximum geometric distortion value of the nominal design is $\sim$ 0.3 $mas$. Assuming a linear combination of the effects, the allowed degradation after the manufacturing, assembly and integration of the instrument is $\sim$ 2.1 $mas$ in order to satisfy the requirement of 2.4 $mas$.  
\section{MAORY prerequisites for the ZeRO}
\label{sec:4}
The MPO integration concept is based on positioning the optics within the accuracy of a LT, then using active compensators with the ZeRO to refine the alignment.\\ 
A dedicated unit for calibration purposes is mandatory to simulate the ELT focal plane in a 180 arcsec FoV at Nasmyth focus of the telescope where MAORY is planned to be placed. This calibration unit will be used to generate a set of reference sources and materialize the ELT focal plane and optical axis. The current calibration unit design is similar to the one which has been developed by NFIRAOS~\cite{lamontagne2017optomechanical}.\\ The PSFs and wavefronts coming from the sources and delivered at the MPO exit focal plane are the observables used for the alignment purposes. Two or more commercial cameras are placed at the MPO focus to detect the PSFs of the sources and some WFSs are used to measure the wavefronts. The WFSs could be commercial sensors dedicated to the the Assembly Integration and Verification (AIV) phase or the instrument NGS WFSs when available. The cameras and WFSs are supposed to be placed on a de-rotator and have to be moved to span the 180 arcsec FoV which is about 600 $mm$ in diameter.
\\The current baseline is to use two LT positions on the bench (see Figure~\ref{fig:1}) in order to integrate all the MAORY opto-mechanical elements. The alignment routine, presented in the following, is then used to restore the nominal optical performances. 
\subsection{Active compensators selection}
As discussed in Section~\ref{sec:2}, the  truncated sensitivity matrix identifies the worst offender DoFs for the system performance. The analytic approach of Section~\ref{sec:2} has been pursued considering different optical parameters, besides the Zernike coefficients, in order to separate and better understand the DoFs perturbations impact on the alignment performance.
The results, listed in the tables below, correspond to the variation of a given optical parameter with respect to the nominal design when moving, one by one, a given DoF of a single optical element. The movements correspond to 1 $mm$ of axial shift or to the corresponding tilt after 1 $mm$ of displacement at the mirror edge. The ratios between the variation and the nominal values are reported since the ratios are more effective in the determination of the DoFs sensitivity to a given parameter. 
\begin{table}[!h]
\caption{After 1 $mm$ DoF perturbation (proper tilt angle to have 1 $mm$ of displacement at the mirror edge) the variation of the mean RMS WFE over the MAORY FoV is considered. The values are the ratio of the variation to the nominal design. The chosen active compensators are in red.}
\label{tab:2}       
\begin{center}  
\begin{tabular}{@{}lllllll@{}}
\toprule
\multicolumn{1}{c}{\textbf{Element}} & \multicolumn{1}{c}{\textbf{$\Delta$x}} & \multicolumn{1}{c}{\textbf{$\Delta$y}} & \multicolumn{1}{c}{\textbf{$\Delta$z}} & \multicolumn{1}{c}{\textbf{$\theta$x}} & \multicolumn{1}{c}{\textbf{$\theta$y}} & \multicolumn{1}{c}{\textbf{$\theta$z}} \\ \midrule
\multicolumn{1}{l|}{\textbf{M6}} & 0.06 & 0.04 & 1.42 & 3.2 & 2.94 & 0.14 \\
\multicolumn{1}{l|}{\textbf{M7}} & 1.16 & 3.36 & 6e-4 & \textcolor{red}{25.78} & \textcolor{red}{22.98} & 8.22 \\
\multicolumn{1}{l|}{\textbf{M8/DM1}} & 0.68 & 0.64 & \textcolor{red}{8.02} & \textcolor{red}{45.72} & \textcolor{red}{44.96} & - \\
\multicolumn{1}{l|}{\textbf{M9/DM2}} & 0.018 & 0.04 & 0.14 & \textcolor{red}{7.22} & \textcolor{red}{8.5} & - \\
\multicolumn{1}{l|}{\textbf{Dichroic}} & - & - & 2.52 & \textcolor{red}{21.6} & \textcolor{red}{20.4} & - \\
\multicolumn{1}{l|}{\textbf{M10}} & 0.1 & 1.02 & 2.52 & 6.88 & 5.52 & 3.04 \\
\multicolumn{1}{l|}{\textbf{M11}} & - & - & 1.42 & 1.1 & 1.34 & -
\end{tabular}
\end{center}  
\end{table}
\newline
\begin{table}[!h]
\caption{Same as Table~\ref{tab:2} but considering the maximum value of the geometric distortion in the MICADO FoV.}
\label{tab:3}       
\begin{center}  
\begin{tabular}{@{}lllllll@{}}
\toprule
\multicolumn{1}{c}{\textbf{Element}} & \multicolumn{1}{c}{\textbf{$\Delta$x}} & \multicolumn{1}{c}{\textbf{$\Delta$y}} & \multicolumn{1}{c}{\textbf{$\Delta$z}} & \multicolumn{1}{c}{\textbf{$\theta$x}} & \multicolumn{1}{c}{\textbf{$\theta$y}} & \multicolumn{1}{c}{\textbf{$\theta$z}} \\ \midrule
\multicolumn{1}{l|}{\textbf{M6}} & 0.017 & 0.033 & 8e-5 & 2.491 & 2.553 & 0.247 \\
\multicolumn{1}{l|}{\textbf{M7}} & 0.170 & 0.230 & 0.008 & 1.902 & 0.975 & 0.690 \\
\multicolumn{1}{l|}{\textbf{M8/DM1}} & 0.214 & 0.287 & 0.002 & \textcolor{red}{9.126} & \textcolor{red}{7.921} & - \\
\multicolumn{1}{l|}{\textbf{M9/DM2}} & 0.022 & 0.136 & 0.008 & \textcolor{red}{7.413} & \textcolor{red}{6.955} & - \\
\multicolumn{1}{l|}{\textbf{Dichroic}} & - & - & 0.0 & \textcolor{red}{5.228} & \textcolor{red}{4.889} & - \\
\multicolumn{1}{l|}{\textbf{M10}} & 0.018 & 0.121 & 0.025 & 1.368 & 1.013 & 0.499 \\
\multicolumn{1}{l|}{\textbf{M11}} & - & - & 0.011 & 0.144 & 0.311 & -
\end{tabular}
\end{center}  
\end{table}
\newline
\begin{table}[!h]
\caption{Same as Table~\ref{tab:2} but considering the exit pupil blur in the MICADO FOV.}
\label{tab:4}       
\begin{center}  
\begin{tabular}{@{}lllllll@{}}
\toprule
\multicolumn{1}{c}{\textbf{Element}} & \multicolumn{1}{c}{\textbf{$\Delta$x}} & \multicolumn{1}{c}{\textbf{$\Delta$y}} & \multicolumn{1}{c}{\textbf{$\Delta$z}} & \multicolumn{1}{c}{\textbf{$\theta$x}} & \multicolumn{1}{c}{\textbf{$\theta$y}} & \multicolumn{1}{c}{\textbf{$\theta$z}} \\ \midrule
\multicolumn{1}{l|}{\textbf{M6}} & 0.0006 & 0.001 & 0.00 & 0.031 & 0.031 & 0.001 \\
\multicolumn{1}{l|}{\textbf{M7}} & 0.001 & 0.002 & 0.00 & 0.043 & 0.026 & 0.031 \\
\multicolumn{1}{l|}{\textbf{M8/DM1}} & 0.002 & 0.004 & 0.00 & \textcolor{red}{0.111} & 0.033 & - \\
\multicolumn{1}{l|}{\textbf{M9/DM2}} & 0.001 & 0.002 & 0.00 & \textcolor{red}{0.094} & 0.035 & - \\
\multicolumn{1}{l|}{\textbf{Dichroic}} & - & - & 0.00 & \textcolor{red}{0.081} & 0.037 & - \\
\multicolumn{1}{l|}{\textbf{M10}} & 0.00 & 0.002 & 0.00 & 0.022 & 0.028 & 0.015 \\
\multicolumn{1}{l|}{\textbf{M11}} & - & - & 0.00 & 0.028 & 0.028 & -
\end{tabular}
\end{center}  
\end{table}
\newline
\begin{table}[!h]
\caption{Same as Table~\ref{tab:2} but considering the exit pupil eccentricity in the MICADO FOV.}
\label{tab:5}       
\begin{center}  
\begin{tabular}{@{}lllllll@{}}
\toprule
\multicolumn{1}{c}{\textbf{Element}} & \multicolumn{1}{c}{\textbf{$\Delta$x}} & \multicolumn{1}{c}{\textbf{$\Delta$y}} & \multicolumn{1}{c}{\textbf{$\Delta$z}} & \multicolumn{1}{c}{\textbf{$\theta$x}} & \multicolumn{1}{c}{\textbf{$\theta$y}} & \multicolumn{1}{c}{\textbf{$\theta$z}} \\ \midrule
\multicolumn{1}{l|}{\textbf{M6}} & 0.00 & 0.013 & 0.004 & 1.108 & 0.002 & 0.001 \\
\multicolumn{1}{l|}{\textbf{M7}} & 0.00 & 0.086 & 0.002 & 0.478 & 0.0431 & 0.004 \\
\multicolumn{1}{l|}{\textbf{M8/DM1}} & 0.00 & 0.108 & 0.008 & \textcolor{red}{3.369} & 0.021 & - \\
\multicolumn{1}{l|}{\textbf{M9/DM2}} & 0.00 & 0.065 & 0.025 & \textcolor{red}{2.717} & 0.008 & - \\
\multicolumn{1}{l|}{\textbf{Dichroic}} & - & - & 0.00 & \textcolor{red}{2.217} & 0.043 & - \\
\multicolumn{1}{l|}{\textbf{M10}} & 0.00 & 0.043 & 0.001 & 0.804 & 0.021 & 0.006 \\
\multicolumn{1}{l|}{\textbf{M11}} & - & - & 0.004 & 0.108 & 0.002 & -
\end{tabular}
\end{center}  
\end{table}
 \newline 
\begin{table}[!h]
\caption{Same as Table~\ref{tab:2} but considering the exit pupil position in the MICADO FOV.}
\label{tab:6}       
\begin{center}  
\begin{tabular}{@{}lllllll@{}}
\toprule
\multicolumn{1}{c}{\textbf{Element}} & \multicolumn{1}{c}{\textbf{$\Delta$x}} & \multicolumn{1}{c}{\textbf{$\Delta$y}} & \multicolumn{1}{c}{\textbf{$\Delta$z}} & \multicolumn{1}{c}{\textbf{$\theta$x}} & \multicolumn{1}{c}{\textbf{$\theta$y}} & \multicolumn{1}{c}{\textbf{$\theta$z}} \\ \midrule
\multicolumn{1}{l|}{\textbf{M6}} & 0.001 & 0.001 & 9e-4 & 0.100 & 0.037 & 0.003 \\
\multicolumn{1}{l|}{\textbf{M7}} & 0.001 & 0.002 & 0.024 & \textcolor{red}{0.197} & 0.037 & 0.038 \\
\multicolumn{1}{l|}{\textbf{M8/DM1}} & 0.002 & 0.006 & 0.001 & \textcolor{red}{0.262} & 0.043 & - \\
\multicolumn{1}{l|}{\textbf{M9/DM2}} & 0.001 & 0.004 & 0.006 & \textcolor{red}{0.217} & 0.041 & - \\
\multicolumn{1}{l|}{\textbf{Dichroic}} & - & - & 0.008 & 0.109 & 0.050 & - \\
\multicolumn{1}{l|}{\textbf{M10}} & 7e-4 & 0.004 & 0.008 & 0.022 & 0.039 & 0.022 \\
\multicolumn{1}{l|}{\textbf{M11}} & - & - & 0.010 & 2e-4 & 0.038 & -
\end{tabular}
\end{center}  
\end{table}
\\The sensitivity analysis shows that RMS WFE (the RSS of Zernike coefficients) is the most sensitive parameter. The worst offenders DoFs are the M8 shift along the optical axis, the tilts of M7, M8, M9 and of the dichroic. They are foreseen to be used as active compensators for the optical alignment. By acting on these DoFs, it is possible to partially compensate the optical parameters variation introduced by the DoFs misalignments not used as active compensators. The worst offender DoFs for the other optical parameters are a subset of the RMS WFE ones. Extending the discussion to the instrument operations (which is beyond the scope of this paper), MAORY needs an active M11 compensation that works in combination with the MCAO architecture (which includes the telescope tracking) to control the exit pupil and focal plane positions.
M11 tilts are then added as compensators during the instrument alignment to relax the control of exit pupil and focal plane position by the others DoFs compensators. The latter will not be active during the operations.
\section{ZeRO ray-tracing simulation}
\label{sec:simulation}
\begin{figure}[!h]
\begin{center}
\begin{tabular}{c}
  \includegraphics[height=5.5cm]{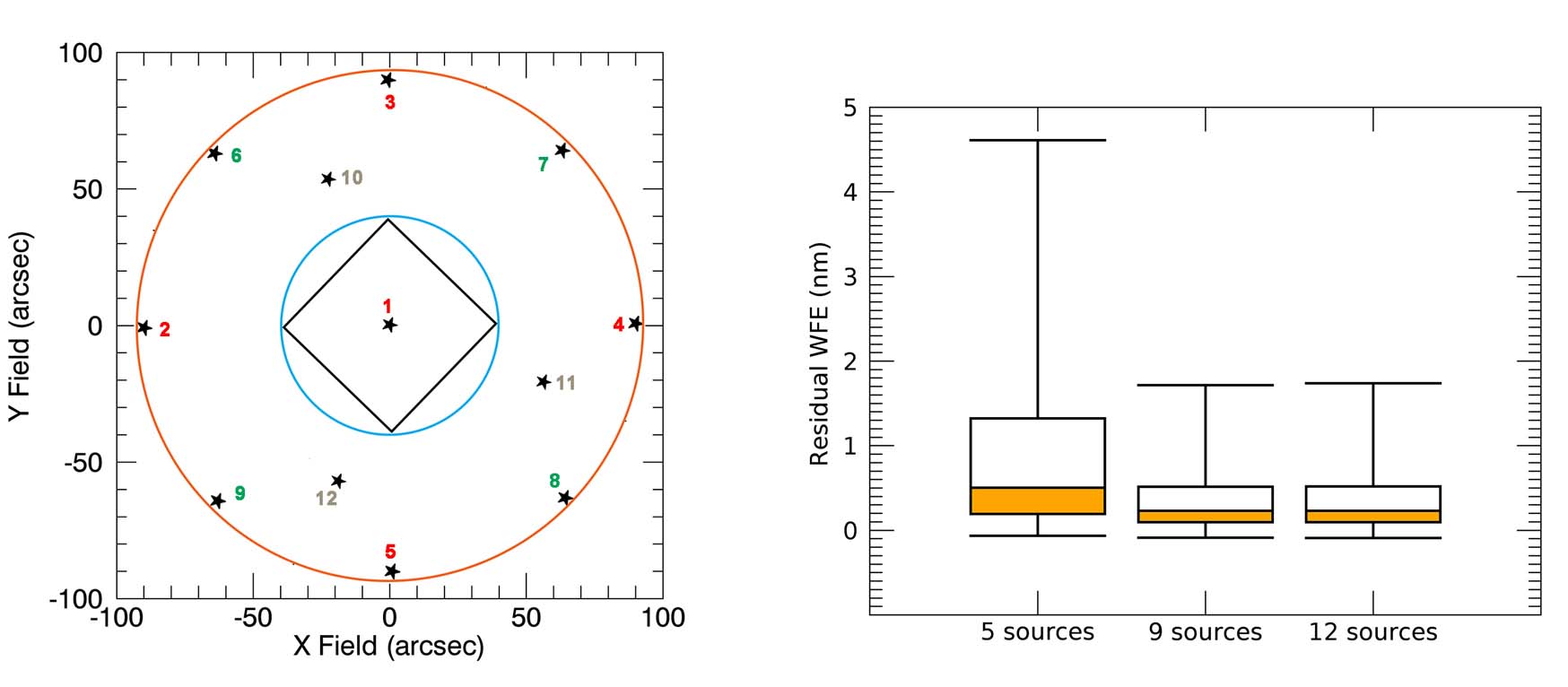}
      \\
  (a) \hspace{5.1cm} (b)
      \end{tabular}
\end{center}
\caption{(a): conceptual layout of the sources at the MAORY entrance focal plane which reproduce the ELT focal plane. (b): Residual RMS WFE as a function of the number of sources (following the geometry shown in (a)). It is the difference between the reference RMS WFE and the measured RMS WFE (mean across the entire FoV) after the alignment procedure. Box plots show minimum and maximum values and quartiles of Monte Carlo trials distribution.}
\label{fig:8}       
\end{figure}

The alignment simulation is based on Zemax Programming Language (ZPL) macros combined with a Monte Carlo approach in order to take into consideration the probabilistic behaviour of the manufacturing and assembly process.
As stated in Section~\ref{sec:2}, the ZeRO alignment algorithm has to refer to an optical design model. The reference design, to which the alignment performance will refer to, is the `as built' or measured design. At the end of the manufacturing process, the shape of the optical surfaces will be measured within a certain precision to verify the compliance to tolerances. The measured design will undergo an optimization of optics positions to take into account the measured low order surface irregularities and mitigate their effects on the MPO optical quality~\cite{patti2018maory}. This optimized design will be the reference. 
In the following simulations the performance degradation after the alignment is referred to the nominal design and not the foreseen `as built' design, which is unknown at the moment of this writing. Since all the presented results are in the form of variation with respect to a reference, the described analysis is independent from the chosen reference design.\\ 
The ray-tracing simulation shall reproduce the main steps of the MPO alignment as the following:
\begin{enumerate}[(a)]
\item	The optics are pre-aligned in their mounts prior to installation on the optical bench structure. 
\item	The LT reads the SMRs locations once the interface plate is installed on the optical bench. Then, the mount is shimmed and settled into position according to the LT measure.
\item	Once the optics are properly located on the bench, they have to be aligned by using the compensators controlled by active components.
\end{enumerate}
Points (a) and (b) are related to the assembly phase of the MPO that is simulated by a random set of assembled optical elements within specified tolerance ranges. For point (a), the considered tolerance range is $\pm$0.2 $mm$ with the assumption that this is the error in measuring the surfaces vertex position (see Table~\ref{tab:9}). As reference, this error is coarser than what prescribed in Chambers et al.~\cite{chambers2002optical}. The used values for point (b) are the predicted LT accuracy, listed in Table~\ref{tab:7}.
\begin{table}[!b]
\caption{Accuracy that characterize the measured optical elements position according to the LT position 1. The accuracy depends on the LT distance from the SMRs and the diameter of the circumscribed circle to the SMRs which define a polygon.}
\label{tab:7}       
\begin{center}  
\begin{tabular}{@{}lcc|ccc@{}}
\toprule
 & \multicolumn{2}{c|}{Laser Tracker} & \multicolumn{3}{c}{Worst case accuracy} \\ \midrule
\multicolumn{1}{l|}{Element} & \textbf{\begin{tabular}[c]{@{}c@{}}distance from\\ SMRs (m)\end{tabular}} & \textbf{\begin{tabular}[c]{@{}c@{}}SMRs\\  Fit circle diameter (m)\end{tabular}} & \textbf{\begin{tabular}[c]{@{}c@{}}Focus\\   ($\mu$m)\end{tabular}} & \textbf{\begin{tabular}[c]{@{}c@{}}Decenter \\ ($\mu$m)\end{tabular}} & \textbf{\begin{tabular}[c]{@{}c@{}}Tip-tilt\\   (arcsec)\end{tabular}} \\ \midrule
\multicolumn{1}{l|}{\textbf{M6}} & 3.8 & 0.91 & 26 & 50 & 14 \\
\multicolumn{1}{l|}{\textbf{M7}} & 0.9 & 0.62 & 25 & 39 & 15 \\
\multicolumn{1}{l|}{\textbf{M8}} & 2.1 & 0.93 & 35 & 35 & 17 \\
\multicolumn{1}{l|}{\textbf{M9}} & 1.8 & 0.93 & 25 & 42 & 13 \\
\multicolumn{1}{l|}{\textbf{Dichroic}} & 2.6 & 0.66 & 38 & 36 & 23 \\
\multicolumn{1}{l|}{\textbf{M10}} & 2.5 & 0.60 & 37 & 36 & 25 \\
\multicolumn{1}{l|}{\textbf{M11}} & 5.7 & 0.66 & 52 & 37 & 32 \\ \bottomrule
\end{tabular}
\end{center}  
\end{table} 
\\500 Monte Carlo trials have been run following a parabolic distribution of the predicted assembly tolerances that we consider as misalignments. Since the parabolic distribution yields selected values that are more likely to be at the extreme ends of the tolerance range, rather than near the middle as for a Gaussian distribution, this approach is conservative.\\ 
The main steps of the ZeRO ray-tracing simulation are:
\begin{enumerate}
\item  Generate $n$ Monte Carlo trials where the optics has been perturbed within the defined ranges of misalignments. 
\item	The shift along the optical axis of M8 and the tilts of M7, M8, M9 and the dichroic are the compensators $\rightarrow C_i$ (variable parameters of the MF)
\item	Zernike coefficients, sources centroids and pupil position of the nominal design are the references $\rightarrow V_i$ (Equation~\ref{eq:5})
\item	Measure 10 Zernike coefficients ($4\leq Z\leq13$) from the SH-WFSs located at different focal points $\rightarrow T_i$ (Equation~\ref{eq:5})
\item	Measure the centroids coordinates from the cameras located at different focal points $\rightarrow T_i$
\item	Measure the exit pupil coordinates of the central field $\rightarrow T_i$
\item	All the measured values $T_i$ are set as targets in the merit function of the reference design $\rightarrow MF$ (Equation~\ref{eq:5})
\item	Optimize the reference design with the Zemax damped least squared algorithm using the compensators as variable parameters.
\item Add M11 tilts as variable compensators to improve the exit pupil and focal plane alignment.
\item	The ZeRO outputs are $C_i$ values that minimize $MF$. These are the motions that should be applied in a reverse way to achieve the MPO alignment.
\end{enumerate}
The method verification has been done assuming that there are no measurement errors or manufacturing uncertainties.\\ 
Field sampling should be considered to address the best trade-off between alignment performance and number of sources. The simulation was run considering three different sets of 5, 9 and 12 sources, arranged as in Figure~\ref{fig:8}, and 500 Monte Carlo trials. The results are summarized in terms of the mean residual RMS WFE. It is the difference of the mean RMS WFE, across the entire FoV, between the reference value and the measured value after the alignment procedure. Boxplots are the distributions of values across the 500 Monte Carlo trials. 
\\The case of 5 sources, placed on a cross geometry, is more critical for the ZeRO accuracy. Going from 9 to 12 sources there are no significant improvements in the ZeRO convergence, thus 9 sources are the best compromise. As highlighted by the sensitivity analysis, the RMS WFE is the leading parameter of the ZeRO validation. If the alignment is successful in terms of RMS WFE, the other optical parameters are within the specification. This statement will be confirmed by the simulation which includes all the considered errors at the end of this paper.
\section{Error analysis}
\label{sec:errors}
Once the optimal number of sources has been defined (eight at the edge of the FoV and one at the centre), the alignment simulation has been run considering different sources of error that are expected to affect the real application of the ZeRO to the MAORY system. These errors are treated separately considering each element involved in the alignment procedure.\\ The observables are extracted from the Monte Carlo trials: point 4 to 6 of the simulation described in Section~\ref{sec:simulation}. The observables are perturbed according to the considered error and used as inputs for the ZeRO: point 7 of Section~\ref{sec:simulation}. The outputs of ZeRo are the compensators movements that should be applied in a reverse way to achieve the MPO alignment. When applied to the Monte Carlo trials, the compensators motions are randomly perturbed in order to consider the motors repeatibility as described in the following sub-section. At the end of this Section, the full simulation, including all errors, shows the expected ZeRO performance in terms of RMS WFE, geometric distortion and main optical parameters variation with respect to the reference design. The simulation also provides the ranges of DoFs motion to achieve a successful alignment.
\subsection{Compensators repeatability}
\begin{figure}[!h]
\begin{center}
\begin{tabular}{c}
  \includegraphics[height=5.5cm]{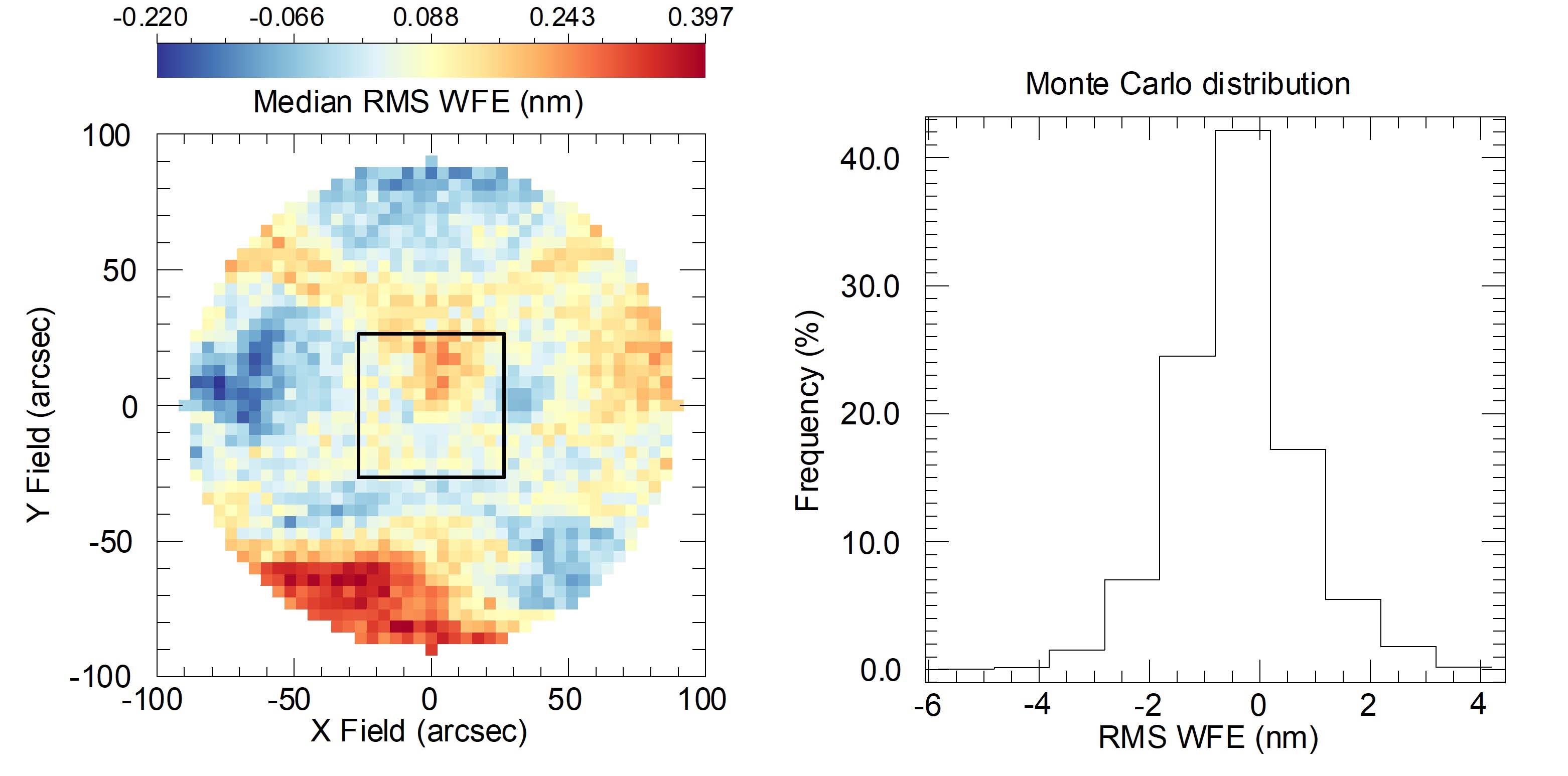}
      \\
  (a) \hspace{5.1cm} (b)
    \end{tabular}
\end{center}
\caption{ZeRO performance simulation in the case of no measurement or manufacturing errors. Variation of the median RMS WFE over the MAORY FoV given by the aligned Monte Carlo trials with respect to the reference design. (a): map of the median variation of the RMS WFE. (b): statistical distribution of the RMS WFE variation across the MAORY FoV within all the Monte Carlo trials. }
\label{fig:6}       
\end{figure}

The ZeRO outputs are given with numerical precision. To care about the active compensators repeatability, the motions are randomly perturbed considering the real motors (available on the market) repeatability which is expected to be about $0.5\:\mu m$.
To be conservative, the simulation considers a repeatability of 2 $\mu m$. Therefore, before applying the DoFs compensations to the Monte Carlo trials, the output values have been randomly perturbed by $\pm$2 $\mu m$.\\
Figure~\ref{fig:6} show the results of the ZeRO in terms of RMS WFE. The plot show the difference between the reference design and the aligned Monte Carlo trials. Median values are very close to zero and the standard deviation is very small ($<$2 $nm$), meaning the alignment has achieved the reference performance.\\
\\After the ZeRO alignment, the worst Monte Carlo trial, in terms of maximum PSF blur, reaches about 0.35 $mas$ at the edge of MICADO FoV which is well below the limit of 2.4 $mas$.
\subsection{Mountings pivot uncertainty}
\label{sec:pivot}	
\begin{figure} [!h]
\begin{center}
\begin{tabular}{c}
  \includegraphics[height=5.5cm]{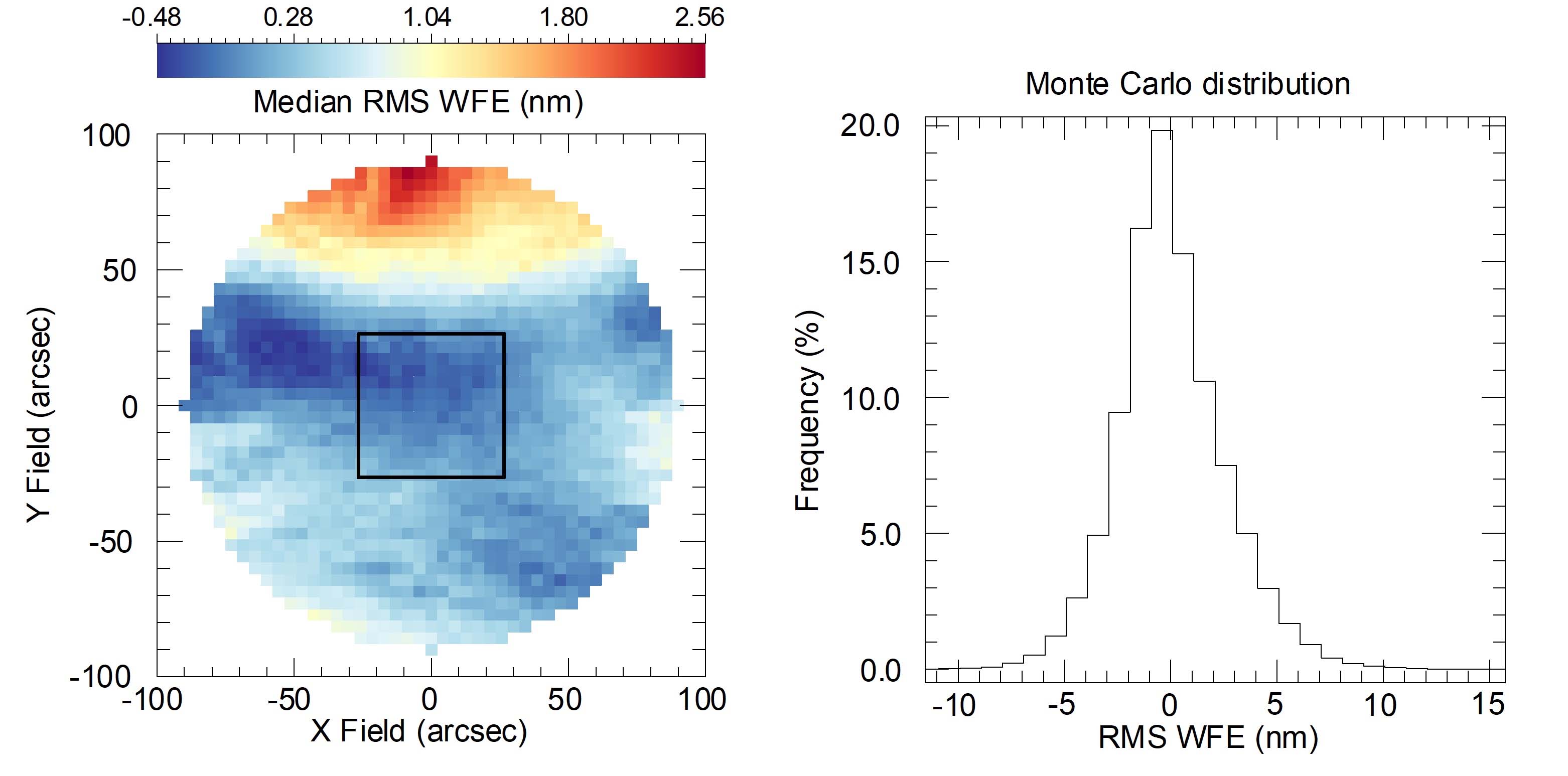}
   \\
  (a) \hspace{5.1cm} (b)
    \end{tabular}
\end{center}
\caption{Same as Figure~\ref{fig:6} but considering an unknown mismatch between the cell pivot point and the optical surface vertex. Within the Monte Carlo trials a random error of $\pm$1.2 $mm$ has been added, in the plane of the optic, between the cell pivot point and the real surface vertex.}
\label{fig:11}       
\end{figure}

The position between the optical element and its mount needs to be characterized before the alignment in order to match the local coordinates of the optical element with the local coordinate of the mount. 
In order to fix the position of the optical element (vertex and axis) with respect to the optical mount, the optical cell is characterized by a coordinate measuring machine (CMM), then the optical element is characterized in the same way.\\ After the characterization of the position of the optical components inside their cells, the next step is the characterization of the axis of rotation around the gimbals. This can be done by a CMM measurement of the cylindrical surface. Then knowing the position of the optical centre with respect to the cell, the optical element is characterized, and using shims or screwing the gimbals it is possible to adjust the position of the rotation axis. This procedure can be done for both the axes; the accuracy of this process depends on the manufacturing process and on the measurement error. If the pivot point of the cell was not centered to the surface vertex, a tilt, to be applied as result of the ZeRO, would become a decentre and an axial shift of the optical element. In particular, a mismatch $d$ between the pivot point and surface vertex, translates into $d\sin\theta$ axial shift and $d(1-\cos\theta)$ decentre, where $\theta$ is the applied tilt angle.\\ A random distance within the range of $\pm$1.2 $mm$ (in the plane of the optic) between the cell pivot point and the real surface vertex has been considered as error. This is a very conservative error which in turn has been used to define the tolerance of mounting. The effect in terms of RMS WFE is quite limited as shown in Figure~\ref{fig:11}.\\ The same effect of the pivot point mismatch is caused by the surface vertex position uncertainty due to manufacturing. For clarity, it has been decided to analyse the errors independently emphasizing a lower accuracy in mechanical integration.\\
\subsection{Errors on calibration unit sources}
\label{sec:sources}	
\begin{figure} [!h]
\begin{center}
\begin{tabular}{c}
  \includegraphics[height=5.5cm]{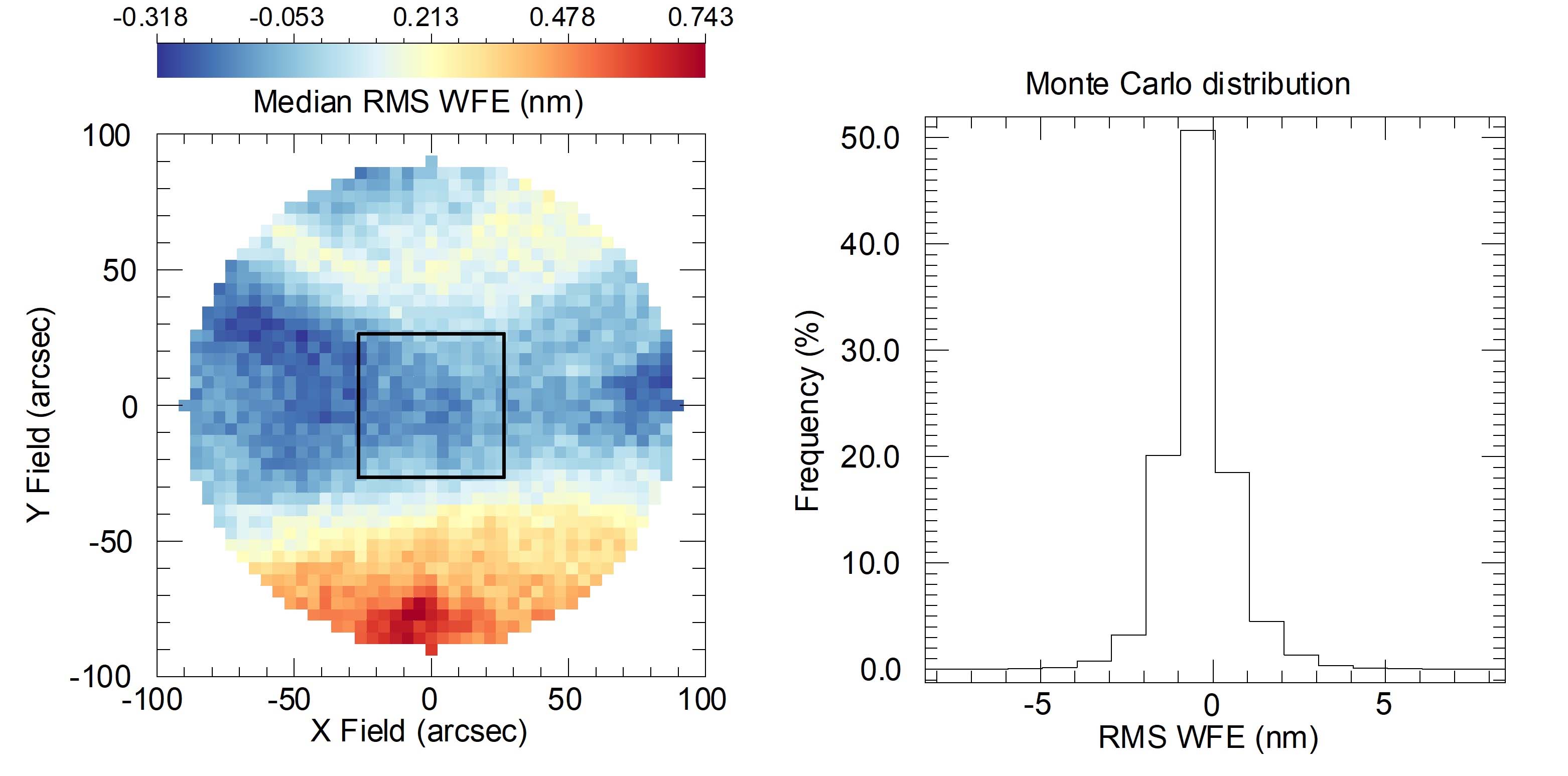}
   \\
  (a) \hspace{5.1cm} (b)
    \end{tabular}
\end{center}
\caption{Same as Figure~\ref{fig:6} but considering errors on $x,y,z$ coordinates of sources.  Within the Monte Carlo trials an additional random error in the range of $\pm$0.1 $mm$ has been considered for sources coordinates.}
\label{fig:9}       
\end{figure}

\begin{figure} [!h]
\begin{center}
\begin{tabular}{c}
  \includegraphics[height=5.5cm]{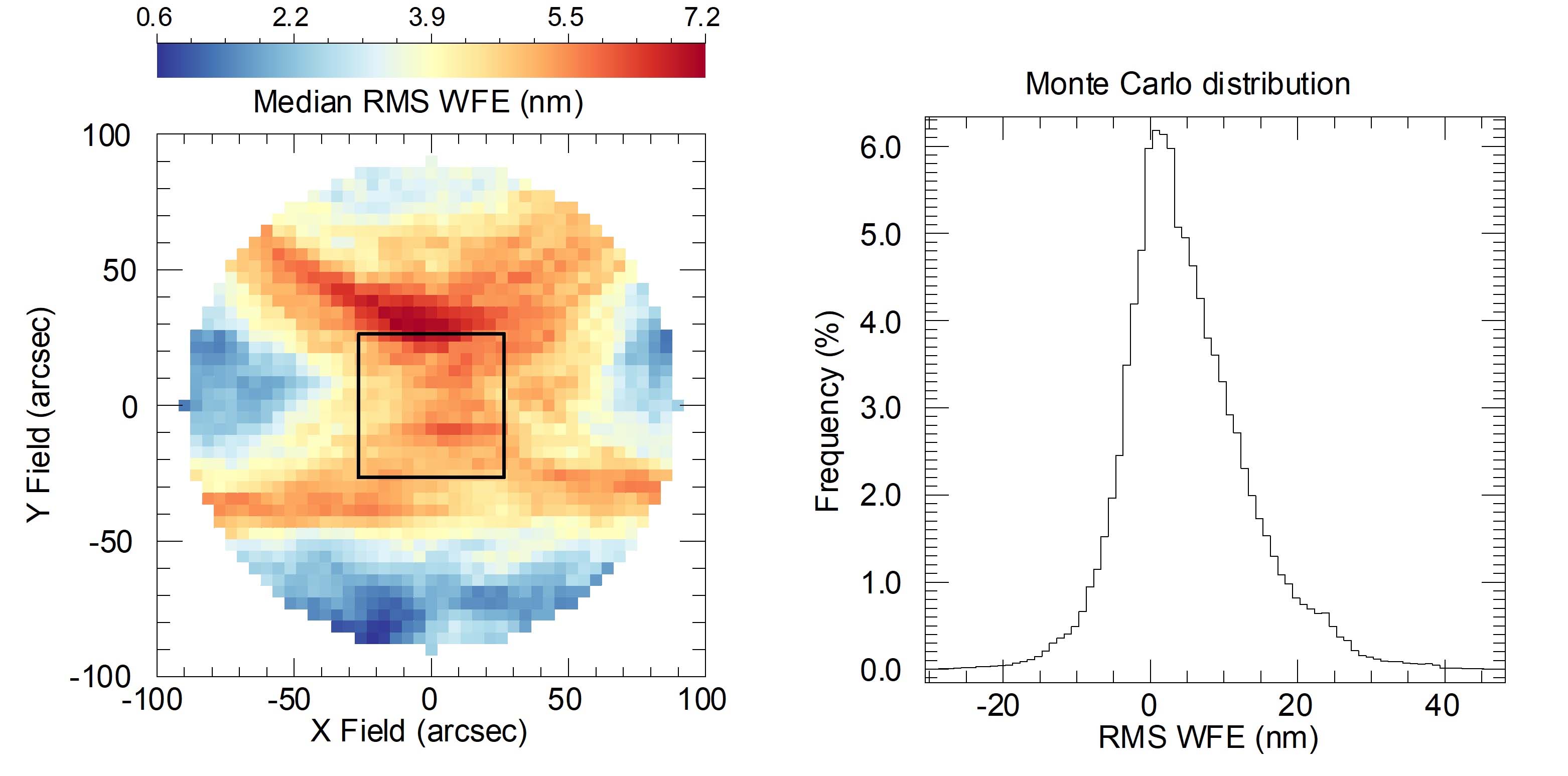}
   \\
  (a) \hspace{5.1cm} (b)
    \end{tabular}
\end{center}
\caption{Same as Figure~\ref{fig:6} but considering errors on sources chief rays angles. Within the Monte Carlo trials an additional random error in the range of $\pm$0.1$^\circ$ has been considered for sources chief rays angles.}
\label{fig:10}       
\end{figure}

To exactly simulate a re-imaged star field by ELT, a set of artificial sources are placed at the entrance focal plane of MAORY. Optical fibres are the baseline choice. The sources must be set with a proper chief ray angle to reproduce the telescope exit pupil position. The surface connecting the sources must be spherical with a curvature radius of 9.88 $m$. The baseline is to implement a limiting aperture in front of each source to define the F-ratio. This solution has been also considered by NFIRAOS~\cite{lamontagne2017optomechanical} where the optical fibre emission angle is limited by conical baffles.\\ The ELT is not telecentric and the chief ray angle of a point-like source changes along the field radius. Thus, at the entrance focal plane, if the sources will be mounted on a circumference, they will have the same chief ray angle and the same sag (the component of the surface displacement from the vertex) reducing the complexity in terms of mechanical manufacturing. \\
The source, which simulates a re-imaged star by ELT at the centre of the FoV, defines the MAORY optical axis and global origin. Therefore, also the exit pupil and focal plane centre positions are defined by the central source. In this context, the considered errors are on the off-axis sources relative to the on-axis one.\\ We do not simulate the intensity profile of the source and/or diffraction effects which could affect the intensity profile of the spots at the image plane and thus, the centroids accuracy (see sub-section~\ref{sec:centroid}).\\
Field errors should be considered as part of the mechanical manufacturing of the mask that simulates the ELT focal plane. Tolerances in the $x,y,z$ coordinates have been allocated as well as tolerance in the beam angle between the surface normal and the incident ray at the surface which defines the ELT focal plane.\\Assumed source position errors are $\pm$0.1 $mm$ and $\pm$0.1$^\circ$ for $x,y,z$ coordinates and chief ray angle, respectively. The effect of these errors is the loss of a direct relation between the observables and the sources defined in the reference design. The worst offender is the $\pm$0.1$^\circ$ chief ray angle variation. Tilting the nominal chief ray angle of a source generates a shift of its beam footprint on the optical surfaces which means a different wavefront and PSF centroid, with respect to the nominal, at the exit focal plane. The same consideration is valid for the $x,y,z$ coordinates variation.
\begin{table}[!h]
\caption{Variation of the observables (Zernike coefficients and centroid) with respect to the nominal values due to the variation of sources parameters. Each value is the mean of the absolute variations across the 9 considered sources.}
\label{tab:8}       
\begin{center}  
\begin{tabular}{@{}ccc@{}}
\toprule
\textbf{}                     & \textbf{Source position ($\pm$0.1 $mm$)} & \textbf{Chief ray angle ($\pm$0.1$^\circ$)} \\ \midrule
\multicolumn{1}{c|}{\textbf{Z4}}       & 0.4 nm                          & 49 nm                  \\
\multicolumn{1}{c|}{\textbf{Z5}}       & 0.01 nm                         & 16 nm                  \\
\multicolumn{1}{c|}{\textbf{Z6}}       & 0.06 nm                         & 39 nm                   \\
\multicolumn{1}{c|}{\textbf{Z7}}       & 0.02 nm                         & 2 nm                      \\
\multicolumn{1}{c|}{\textbf{Z8}}       & 0.01 nm                         & 2 nm                       \\
\multicolumn{1}{c|}{\textbf{Z9}}       & 0.04 nm                         & 2.5 nm                     \\
\multicolumn{1}{c|}{\textbf{Z10}}      & 0.0 nm                          & 2.3 nm                      \\
\multicolumn{1}{c|}{\textbf{Z11}}      & 0.01 nm                         & 0.9 nm                      \\
\multicolumn{1}{c|}{\textbf{Z12}}      & 0.0 nm                          & 1.4 nm                        \\
\multicolumn{1}{c|}{\textbf{Z13}}      & 0.0 nm                          & 0.5 nm                         \\ 
\multicolumn{1}{c|}{\textbf{Centroid}} & 0.1 mm                           & 0.001 mm                   \\   \bottomrule
\end{tabular}
\end{center}  
\end{table}
\\The effect on the observables due to the variation of sources parameters are summarized in Table~\ref{tab:8} in terms of variation with respect to the nominal values. This means the ZeRO simulation was run with that bias on the observables.
\\The sources shall deliver the proper focal aperture (F/17.74). An error on this parameter produces two effects: 
\begin{enumerate}
\item	A different wavefront at the exit focal plane since the beam footprints on the optics are different with respect to the reference case even though the chief rays are coincident. A relative error of 1$\%$ translates to a variation of the reference wavefront of about 2-3 $nm$ RMS in the worst case. 
\item	A different response of the WFS since the pupil size on the lenslet array changes its size. 
At the present stage both errors have been considered to be inside the WFS error budget described in Section~\ref{sec:wfs}. 
\end{enumerate}
\pagebreak
The results, in terms of RMS WFE, of the ZeRO simulation considering errors on $x,y,z$ coordinates of sources and chief ray angles are shown respectively in Figure~\ref{fig:9} and Figure~\ref{fig:10}.
\subsection{Optics manufacturing uncertainty}
\label{sec:manufacturing}
\begin{figure}[!h]
\begin{center}
\begin{tabular}{c}
  \includegraphics[height=5cm]{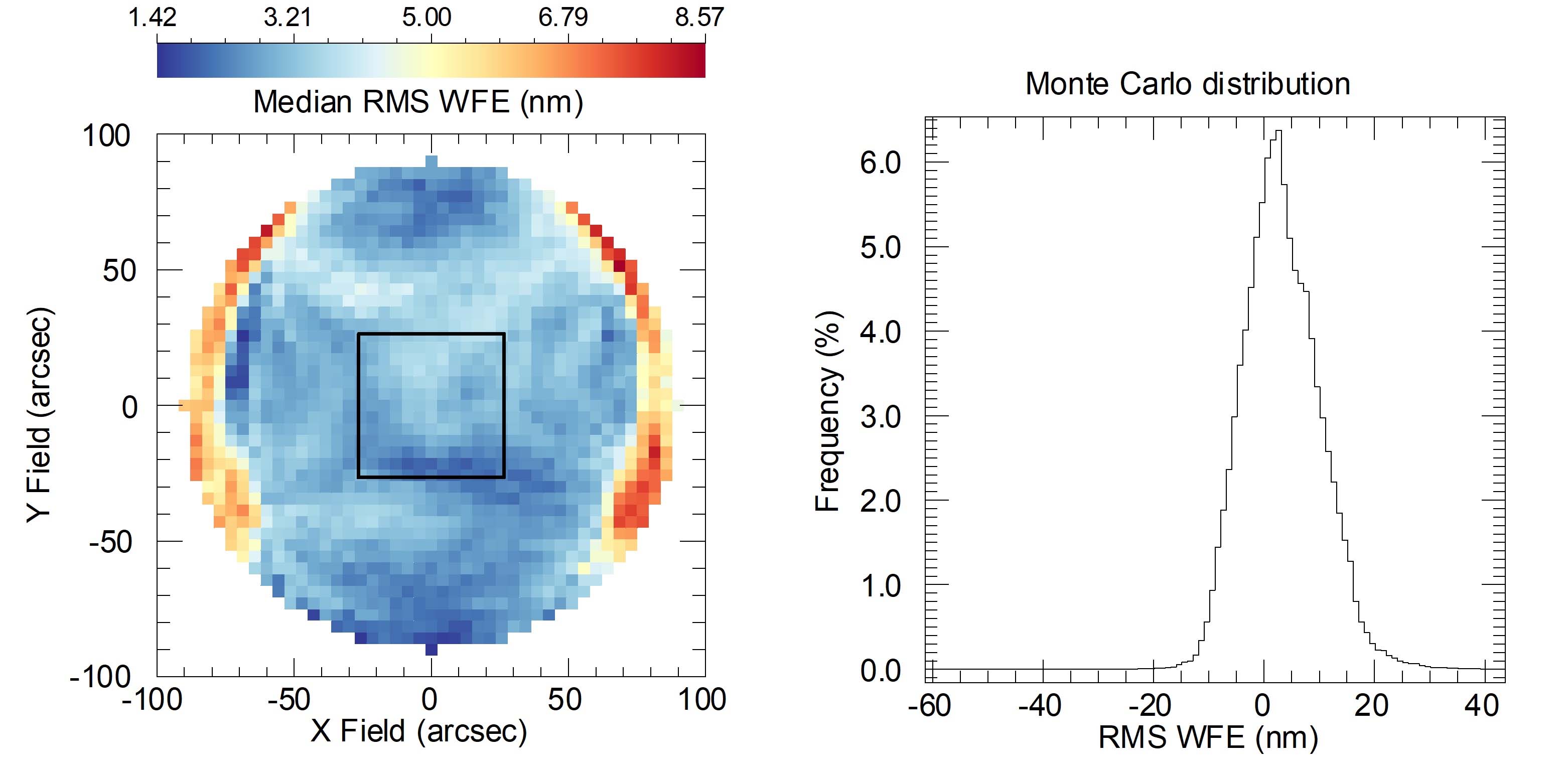}
   \\
  (a) \hspace{5.1cm} (b)
    \end{tabular}
\end{center}
\caption{Same as Figure~\ref{fig:6} but considering unknown LO irregularities of optical surface as listed in Table~\ref{tab:9}. Random values in the ranges of Table~\ref{tab:9} (that considers LO) have been added to optical surfaces within the Monte Carlo trials.}
\label{fig:12}       
\end{figure}

\begin{figure}[!h]
\begin{center}
\begin{tabular}{c}
  \includegraphics[height=5cm]{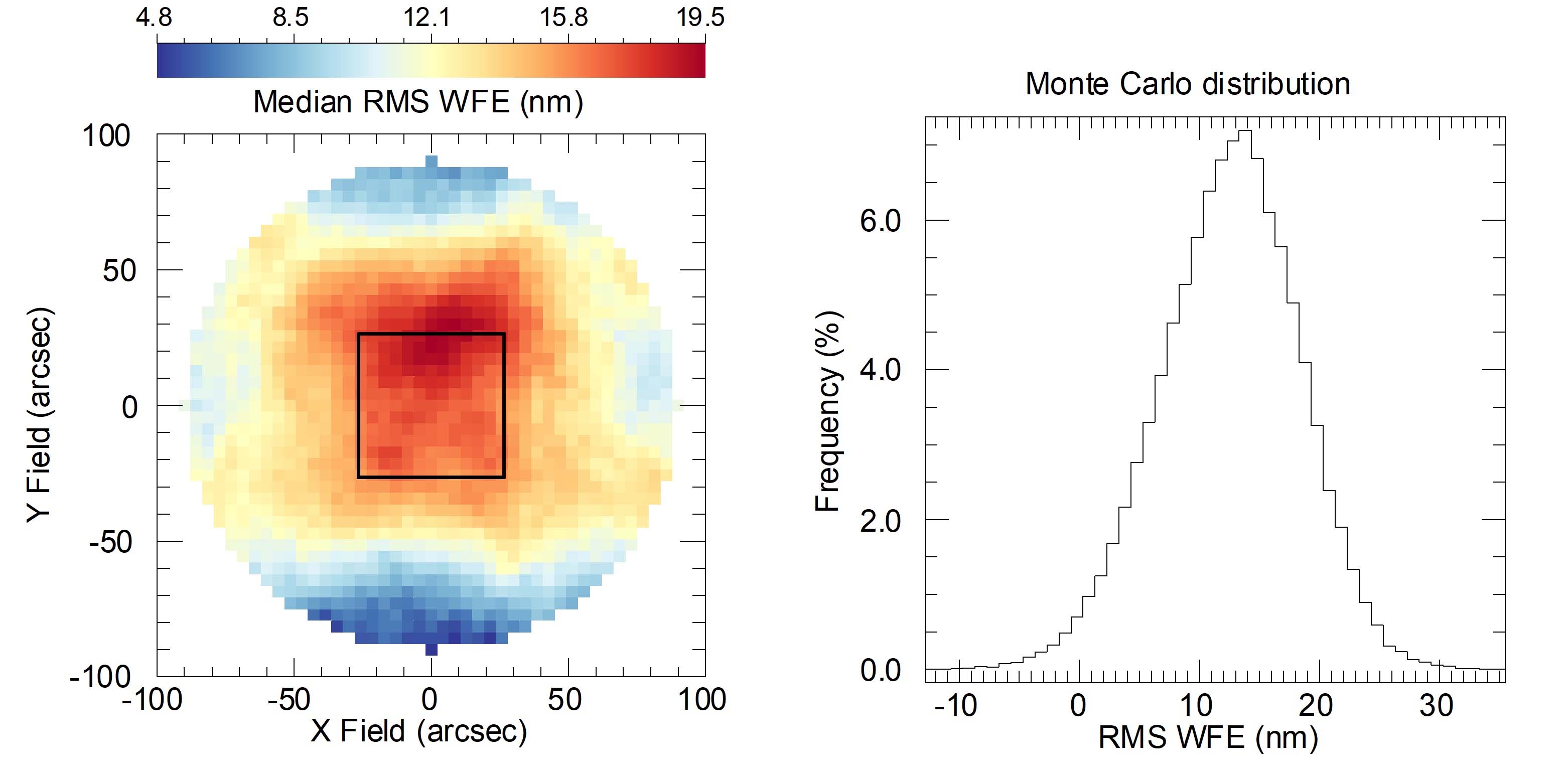}
   \\
  (a) \hspace{5.1cm} (b)
    \end{tabular}
\end{center}
\caption{Same as Figure~\ref{fig:6} but considering unknown HO irregularities of optical surface as listed in Table~\ref{tab:9}. Random values in the ranges of Table~\ref{tab:9} (that considers HO) have been added to optical surfaces within the Monte Carlo trials.}
\label{fig:13}       
\end{figure}

\begin{figure}[!h]
\begin{center}
\begin{tabular}{c}
  \includegraphics[height=5.1cm]{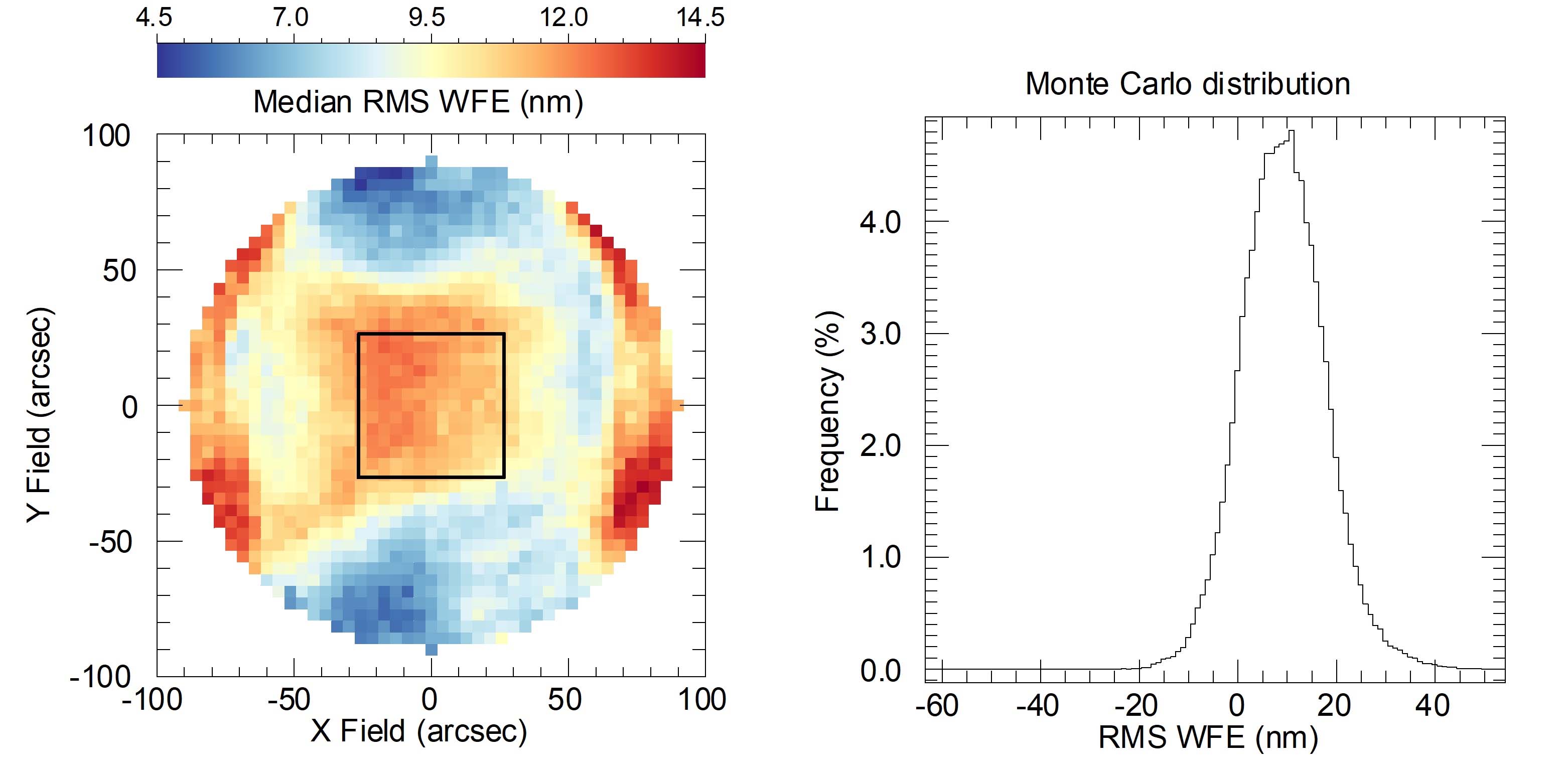}
   \\
  (a) \hspace{5.1cm} (b)
    \end{tabular}
\end{center}
\caption{Same as Figure~\ref{fig:6} but considering unknown LO and HO irregularities of optical surface as listed in Table~\ref{tab:9}. Random values in the ranges of Table~\ref{tab:9} have been added to optical surfaces within the Monte Carlo trials.}
\label{fig:14}       
\end{figure}

The mirrors are assumed to be delivered by the suppliers with a full characterization of their surfaces. In principle, the measured values of surface parameters are implemented in the nominal design which becomes the `as-built' reference design. In case the information is missing or inaccurate, the impact of surface irregularities on the effectiveness of the algorithm is shown in Figure~\ref{fig:14}.
The irregularities have been divided into Low-Order (LO) and High-Order (HO) and modelled by Zernike polynomials~\cite{noll1976zernike}. Table~\ref{tab:9} lists the manufacturing uncertainties used in the simulation.\\ 500 Monte Carlo trials have been run where, in addition to the assembly tolerances, all the manufacturing uncertainties of Table~\ref{tab:9} have been used to perturb the optical surfaces.
\begin{table}[!h]
\caption{Expected uncertainties regarding the optical manufacturing process. Zernike values are specified by the tolerance analysis. Surface irregularities (deviations from the nominal surface) are modelled by means of standard Zernike coefficients whose max tolerance value is the exact RMS error of the surface. When a range of modes is indicated, the value refers to the RSS of values within that range.}
\label{tab:9}       
\begin{center}  
\begin{tabular}{@{}cccccccc@{}}
\cmidrule(l){3-8}
 & \multicolumn{1}{c|}{} & \multicolumn{3}{c|}{\textbf{\begin{tabular}[c]{@{}c@{}}Low orders\end{tabular}}} & \multicolumn{3}{c}{\textbf{\begin{tabular}[c]{@{}c@{}}High orders\end{tabular}}} \\ \midrule
\multicolumn{1}{l}{Element} & \begin{tabular}[c]{@{}c@{}}Vertex\\  position\\ (mm)\end{tabular} & \begin{tabular}[c]{@{}c@{}}Radius of \\ curvature\end{tabular} & \begin{tabular}[c]{@{}c@{}}Zern \\ 5 to 10\end{tabular} & \begin{tabular}[c]{@{}c@{}}Zern \\ 11 \end{tabular} & \begin{tabular}[c]{@{}c@{}}Zern \\ 12 to 28\end{tabular} & \begin{tabular}[c]{@{}c@{}}Zern \\ 29 to 45\end{tabular} & \begin{tabular}[c]{@{}c@{}}Zern \\ 46 to 55\end{tabular} \\ \midrule
\multicolumn{1}{c|}{\textbf{M6}} & $\pm$0.2 & $\pm$0.02\% & 24 nm & 5 nm & 5 nm & 5 nm & 5 nm \\
\multicolumn{1}{c|}{\textbf{M7}} & $\pm$0.2 & $\pm$0.02\% & 18 nm & 5 nm & 5 nm & 3 nm & 5 nm \\
\multicolumn{1}{c|}{\textbf{M8/DM1}} & $\pm$0.2 & $\pm$0.02\% & 12 nm & 5 nm & 5 nm & 5 nm & 5 nm \\
\multicolumn{1}{c|}{\textbf{M9/DM2}} & $\pm$0.2 & $\pm$0.02\% & 12 nm & 5 nm & 5 nm & 3 nm & 5 nm \\
\multicolumn{1}{c|}{\textbf{Dichroic}} & - & $\pm$0.5 fr & 11 nm & 5 nm & 5 nm & 5 nm & 5 nm \\
\multicolumn{1}{c|}{\textbf{M10}} & $\pm$0.2 & $\pm$0.02\% & 20 nm & 5 nm & 5 nm & 3 nm & 5 nm \\
\multicolumn{1}{c|}{\textbf{M11}} & - & $\pm$0.5 fr & 23 nm & 5 nm & 5 nm & 5 nm & 5 nm \\ \bottomrule
\end{tabular}
\end{center}  
\end{table}
\\There is not a specific of manufacturing process that drives the definition of Zernike terms into ranges. In Zemax simulation, one can define a range of Zernike coefficients and an RMS value (Zemax operand 'TEZI'). The optical surface is modelled by the Zernike polynomials whose coefficients are set to a value so that the square root of the sum of the squares of the coefficients yields the specified RMS value. For the Monte Carlo analysis each polynomial term is assigned a coefficient randomly chosen in the range $\pm$1, and the resulting coefficients are then normalized to yield the exact RMS tolerance. When considering a wide range of Zernike coefficients, one has to deal with the power spectra of the coefficients within the range. Zemax satisfies the Monte Carlo statistics but could generate surfaces with larger RMS irregularities in the highest edge of the Zernike coefficients range. In the tolerance analysis~\cite{patti2018maory}, we wanted to avoid this kind of bias and a sub-division into ranges can give us more control on the statistic. The number of Zernike coefficients we can simulate is not infinite, so we stopped to Zernike 55 to speed up the simulation (a lot of coefficients require a lot of simulation time). However, the values of the last Zernike range could be interpreted as Z$>$45 for the purpose of a manufacturing process.
\\Zernike coefficients values (RMS error of the surface) are the same as the MAORY tolerance analysis specifications~\cite{patti2018maory}. The available DoFs are able to partially compensate aberrations until Zernike coefficient 11 (the `spherical' term), hence the computed LO irregularities include Zernike polynomials from 4 (the defocus term) to 11. The uncertainty about the radius of curvature is one order of magnitude lower than its specified tolerance and it is a reasonable expected measurement accuracy~\cite{gerchman1980differential}~\cite{hao2017vertex}.\\ Figure~\ref{fig:12} and Figure~\ref{fig:13} show the separate contribution in term of RMS WFE variation after the ZeRO affected by LO and HO surface uncertainties, respectively. Figure~\ref{fig:14} is the result of ZeRO when both LO and HO irregularities are considered.
\subsection{WFS measurement accuracy}
\label{sec:wfs}	
The MAORY WFSs are supposed to work in closed loop and they are requested to provide relative (differential) wavefront measurements, rather than absolute ones. Nevertheless, after calibrations they can give fairly accurate wavefront measurements. By inserting a point source at the entrance focal plane of the WFS, it is possible to measure the internal systematic WFE due to the WFS itself and use it as slope offsets. The exit pupil displacement can be corrected by internal pupil adjustment in the WFS, e.g. by tilting a pick-off mirror. When measuring an off-axis wavefront, the WFS can be centered by nulling the tilt signal and the pupil centering can be adjusted if required. In this way it is assumed that the WFE measurement are in the same conditions as the on-axis calibration and that a reasonable measurement error can be considered. Taking into account the calibrations and the measurement strategy of the wavefront, it is assumed that the WFS has 10 $nm$ RMS per mode of measurement accuracy. We have not considered a detailed error break-down for the wavefront sensor accuracy. 10 $nm$ RMS per mode is a random error which has been added to the observables extracted from the Monte Carlo trials: point 4 of the simulation described in Section~\ref{sec:simulation}. This assumption is supported by the declared accuracy of most commercial WFSs.
\begin{figure}[!ht]
\begin{center}
\begin{tabular}{c}
  \includegraphics[height=5.5cm]{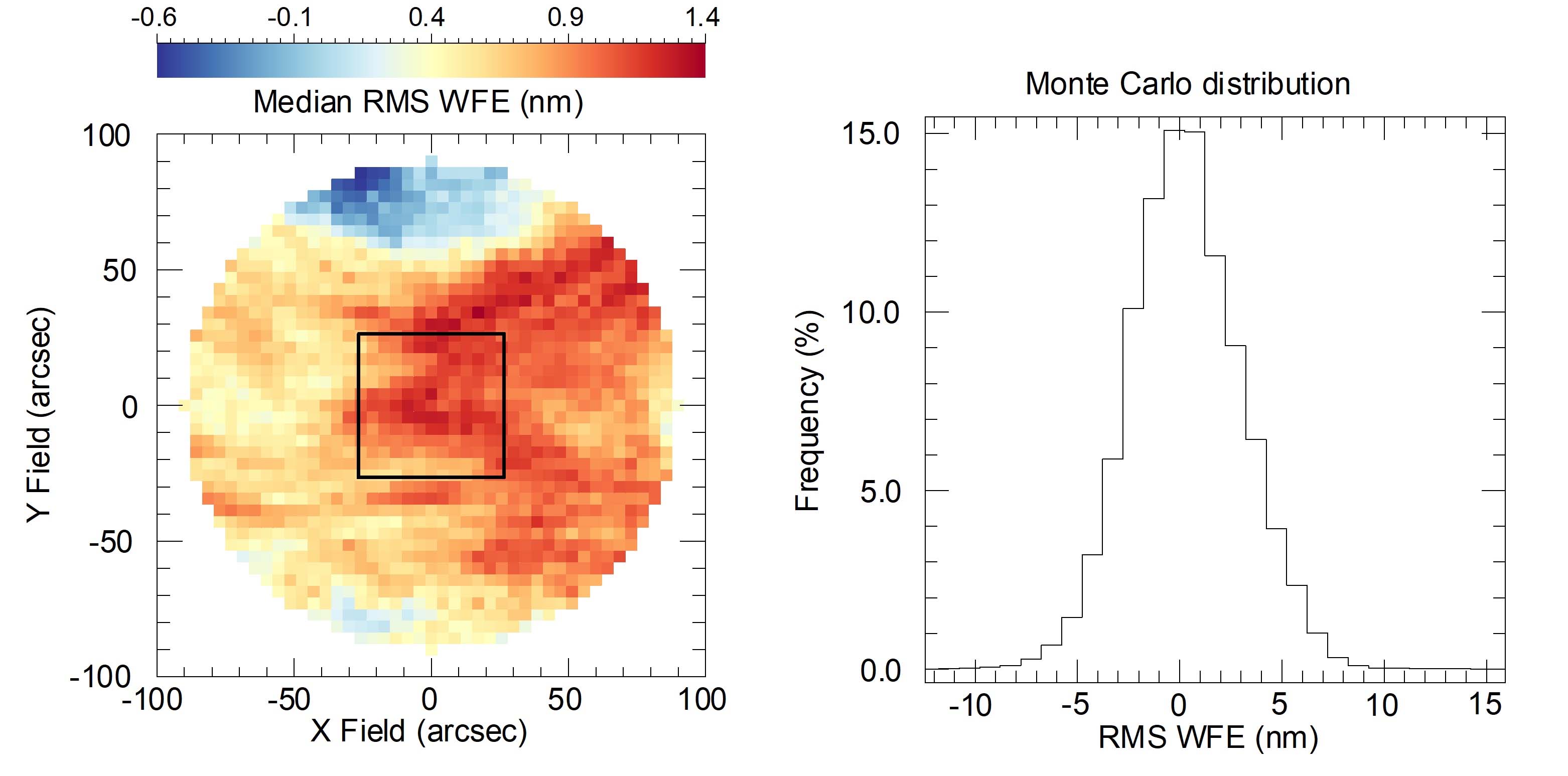}
   \\
  (a) \hspace{5.1cm} (b)
    \end{tabular}
\end{center}
\caption{Same as Figure~\ref{fig:6} but considering errors in the wavefront measurements. Within the Monte Carlo trials a random error of 10 $nm$ RMS per mode has been added to the wavefront measurements.}
\label{fig:15}       
\end{figure}
\subsection{Centroid measurement accuracy}
\label{sec:centroid}	
\begin{figure}[!h]
\begin{center}
\begin{tabular}{c}
  \includegraphics[height=5.5cm]{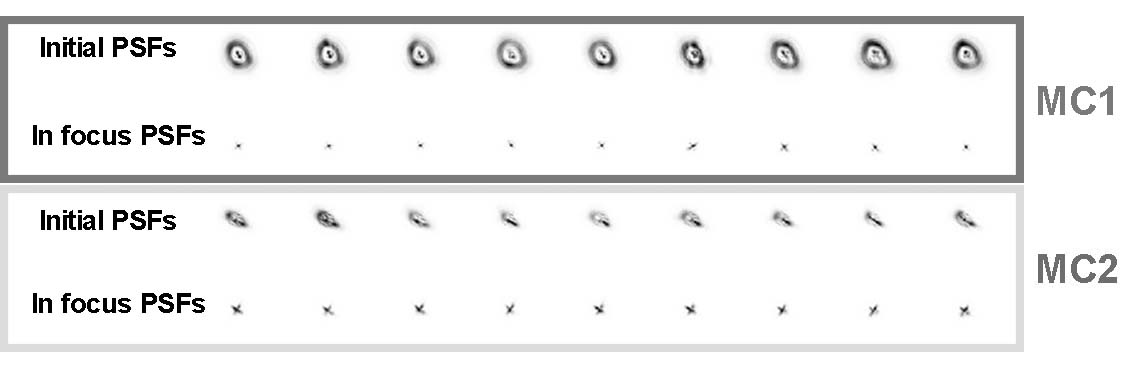}
    \end{tabular}
\end{center}
\caption{Two samples of PSFs taken from two Monte Carlo trials where it is evident the spot shrinking after removing the defocus term.}
\label{fig:17}       
\end{figure}

\begin{figure} [!t]
\begin{center}
\begin{tabular}{c}
 \includegraphics[height=5.5cm]{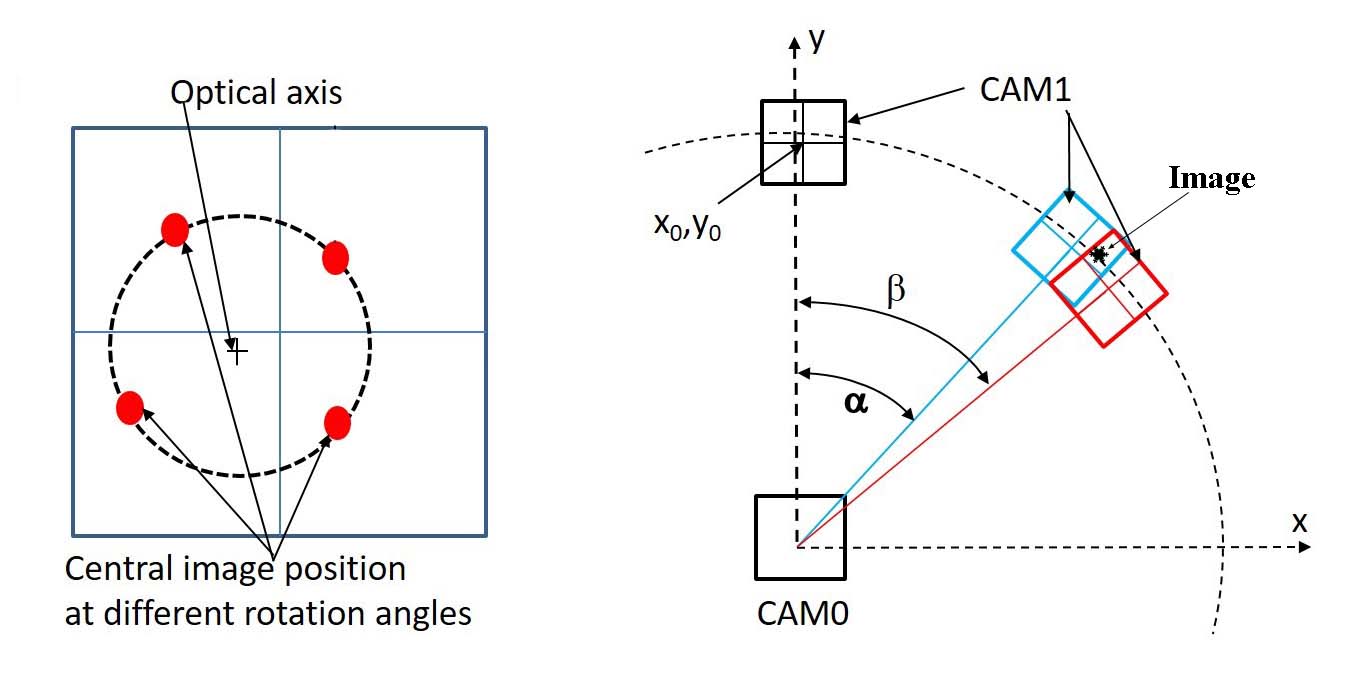}
    \\
  (a) \hspace{5.1cm} (b)
    \end{tabular}
\end{center}
\caption{(a): central image movement as seen by the camera local reference coordinates and at different de-rotator angles in case of displacement with respect to the optical axis. (b): sketch describing the calculation of CAM1 absolute position using the data from two different rotation angles.}
\label{fig:16}       
\end{figure}

The centroids are used to align the exit focal plane with the de-rotator mechanical axis. The measurement accuracy of the centroids depends on several factors: positional accuracy of the measurement camera, PSF shapes and illumination profiles of the sources. With artificial sources, the detectors work in the photon noise regime.\\ After the LT alignment the dominant aberration is defocus. The effect of defocus is the enlargement of the PSFs which could cause the loss of accuracy in the centroid measurement. Figure~\ref{fig:17} shows the 9 PSFs of two different Monte Carlo trials before and after the removal of the defocus term. This could be done by shifting the exit focal plane, or the entrance focal plane, or moving M11 or M8 along the optical axis. M8 is the only compensator for focus among the DoFs used by the ZeRO. The spot shrinking is evident. In the worst case the radius of the out-of-focus PSF, at which 90\% of the energy is encircled, is about 300 $\mu m$. Correcting the defocus, the radius decreases to about 50 $\mu m$.  In the ZeRO simulation the centroids are measured on the PSFs delivered by Zemax. In the real system the centroids coordinates could be different due to optical fibre effects or diffraction effects at the calibration unit level. We assume that any systematic error due to an ‘irregular intensity profile’, coming from the optical fibre and/or diffraction effects, is within 5 $\mu m$ which is (as a rule of thumb) 1/10 of the worst expected spot radius.
\\The measurement of the absolute value of the off-axis image positions, at the exit focal plane, is given by the determination of the off-axis camera positions with respect to the optical axis. The mechanical axis of the de-rotator defines the optical axis which has to be coincident with the image position of the central source. At the central camera local coordinates (CAM0), when the system is not aligned, the PSF moves on a circle centered to the optical axis. It is possible to calculate the absolute position of the PSF which is given by the circle radius (see Figure~\ref{fig:16}).\\Regarding the image position of the off-axis camera (CAM1), it is possible to measure the radial distance of CAM1 by using the de-rotator. Let us consider the coordinates $x_0,\:y_0$ of CAM1 where the mechanical axis of the de-rotator is the origin. After rotation by an angle $\beta$, the camera detects the source image whose coordinates are:
\begin{equation}
\begin{split}
x_s=x_{0,\beta}+x_{s,\beta} \\ 
y_s=y_{0,\beta}+y_{s,\beta} 
\end{split}
\end{equation}
$x_{0,\beta},\: y_{0,\beta}$ are the new camera coordinates after the rotational transformation and $x_{s,\beta},\: y_{s,\beta}$ are the spot centroid coordinates with respect to the camera local reference system. These centroid coordinates can be brought to the camera local reference system before the rotation by using the transformation:
\begin{equation}
\begin{split}
\Delta x_\beta= x_{s,\beta} \cos⁡\beta+y_{s,\beta}\sin\beta \\
\Delta y_\beta= y_{s,\beta} \cos⁡\beta-x_{s,\beta}\sin\beta
\end{split}
\end{equation}
If the same process was done with an angle $\alpha\neq\beta$, it would be possible to measure $\Delta x_\alpha$ and $\Delta y_\alpha$. Defining:
\begin{equation}
\begin{split}
\Delta x=\Delta x_\alpha-\Delta x_\beta \\
\Delta y=\Delta x_\alpha-\Delta x_\beta 
\end{split}
\end{equation}
After some algebra, the $x_0,y_0$ coordinates of CAM1 are:
\begin{equation} 
\begin{split}
x_0=\Delta x\: \frac{(\cos\beta⁡-\cos⁡\alpha)}{2∙[1-\cos⁡(\alpha-\beta) ] }-\Delta y\: \frac{(\sin⁡\beta-\sin⁡\alpha)}{2∙[1-\cos⁡(\alpha-\beta) ] } \\ 
y_0=\Delta y\: \frac{(\cos\beta⁡-\cos⁡\alpha)}{2∙[1-\cos⁡(\alpha-\beta) ] }+\Delta x\: \frac{(\sin⁡\beta-\sin⁡\alpha)}{2∙[1-\cos⁡(\alpha-\beta) ] }
\end{split}
\end{equation}
\\This is the position of CAM1 with respect to the optical axis necessary to know the absolute positions of the PSF centroids.
The worst offender of these measurements could be the de-rotator accuracy, which implies the uncertainty about the knowledge of $\alpha$ and $\beta$ angles. We assume it would be possible to measure the rotational angle by using the LT and thus, the off-axis centroids should be known with an error of $\approx$50 $\mu m$ which is the accuracy level of the LT. The alignment goal is to center the optical axis, defined by the central source, with the de-rotator axis. In principle, the absolute central image position is sufficient to achieve the goal and it is not affected by de-rotator errors since is given by the radius of a circle as previosly described. The ZeRO uses the off-axis centroids to better constrain the MF but the weighting value of these centroids is 100 times lower. During the ZeRO simulation 55 $\mu m$ has been allocated as centroid accuracy which correspond to 16.5 $mas$ on sky. To be conservative, this is the order of magnitude accuracy of the LT plus any possible effects coming from the artificial sources and corresponds to about the size of the worst in-focus PSF, as discussed above.
\subsection{All errors simulation}
\label{sec:all}
\begin{figure}[!h] 
\begin{center}
\begin{tabular}{c}
  \includegraphics[height=5.5cm]{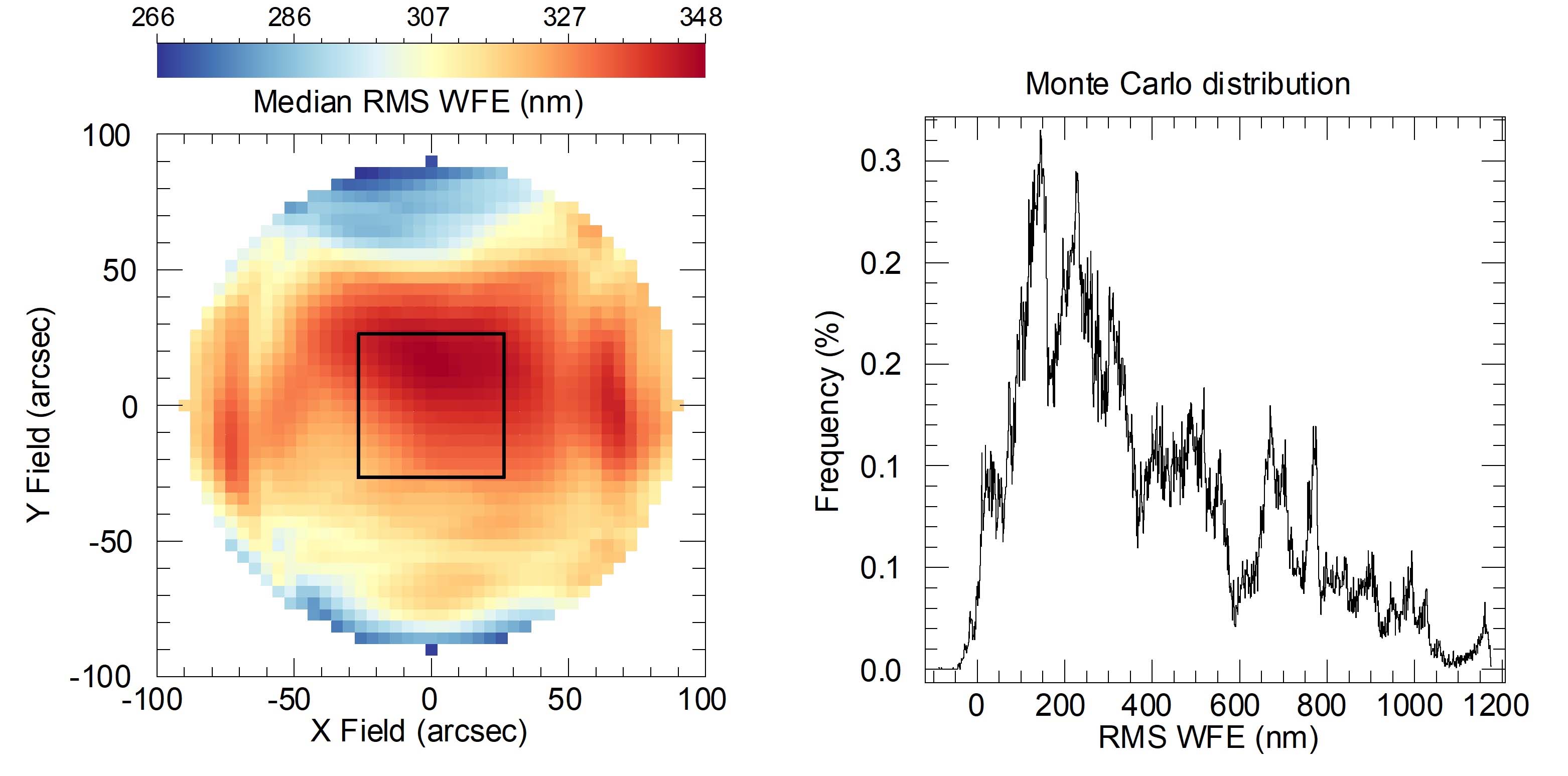}
      \\
  (a) \hspace{5.1cm} (b)
    \end{tabular}
\end{center}
\caption{(b) RMS WFE distribution of the Monte Carlo trials and its median map (a) at the beginning of the procedure, when ZeRO has not achieved the alignment condition yet. This includes all the considered errors.}
\label{fig:18}       
\end{figure}

\begin{figure} [!ht] 
\begin{center}
\begin{tabular}{c}
  \includegraphics[height=5.5cm]{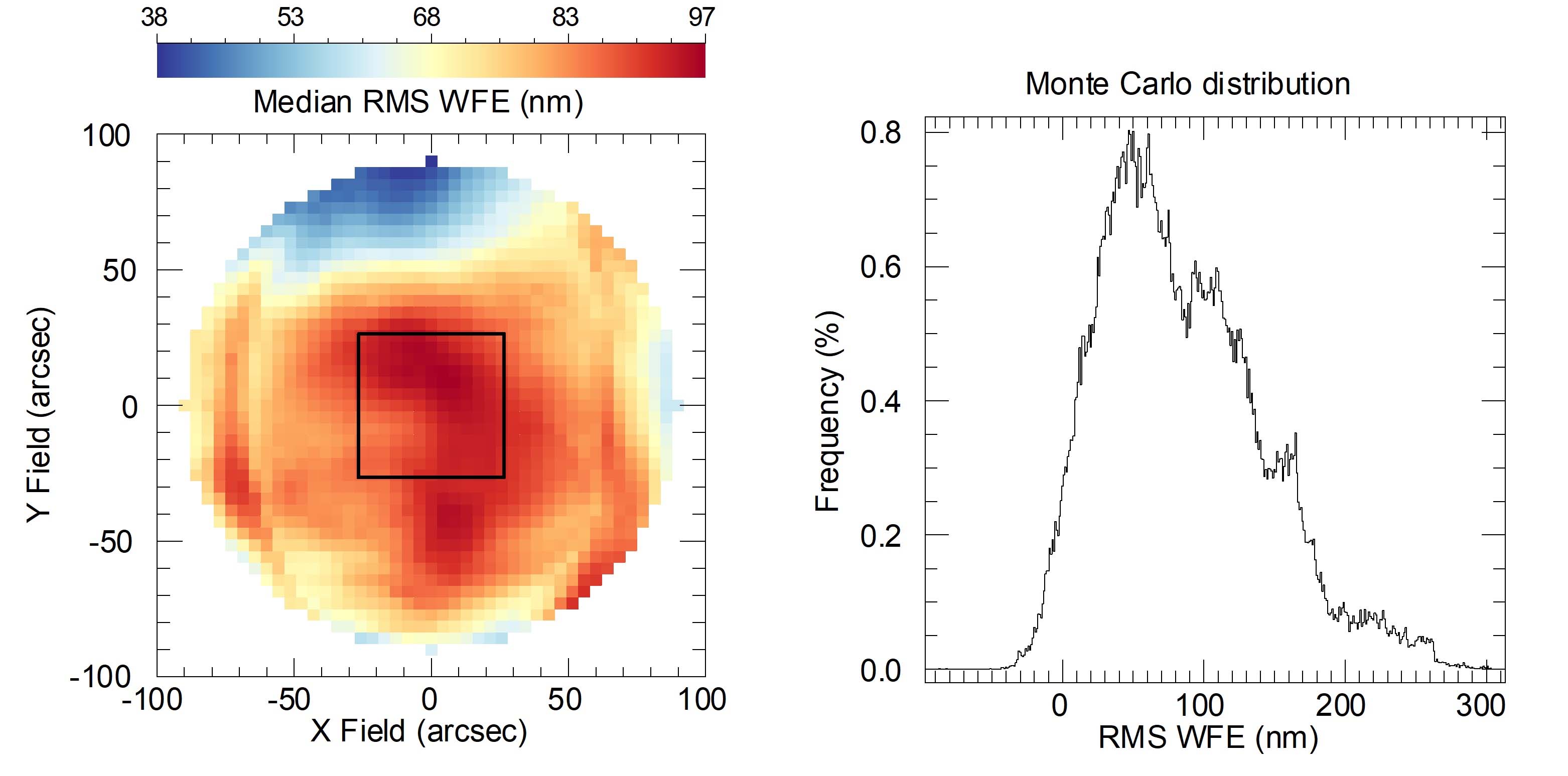}
   \\
  (a) \hspace{5.1cm} (b)
    \end{tabular}
\end{center}
\caption{Same as Figure~\ref{fig:18} but removing the mean defocus contribution to the RMS WFE.}
\label{fig:19}       
\end{figure}

\begin{figure} [!h] 
\begin{center}
\begin{tabular}{c}
  \includegraphics[height=5.5cm]{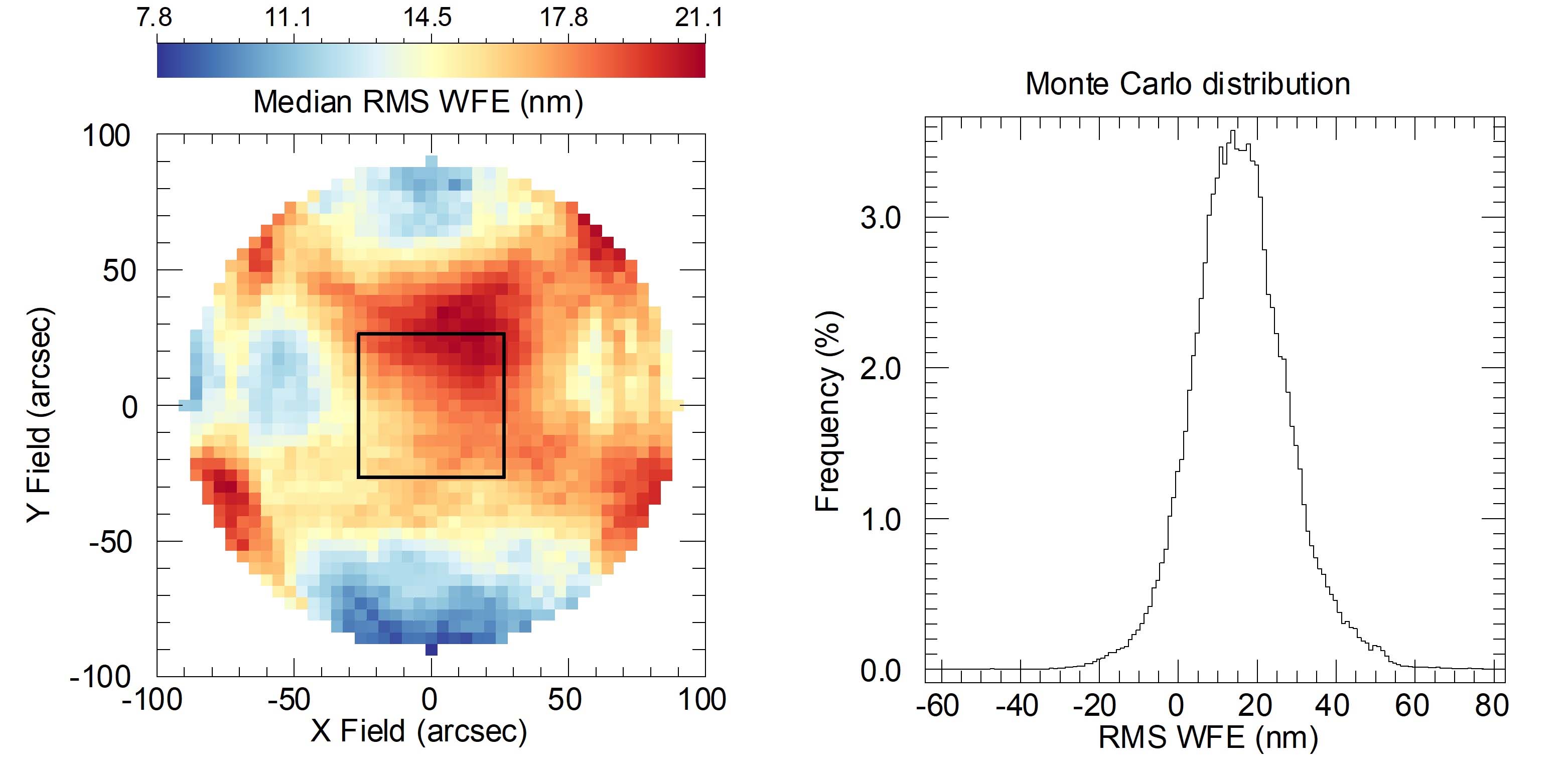}
   \\
  (a) \hspace{5.1cm} (b)
    \end{tabular}
\end{center}
\caption{Same as Figure~\ref{fig:6} but considering the full errors simulation.}
\label{fig:20}       
\end{figure}

\begin{figure} [!h] 
\begin{center}
\begin{tabular}{c}
  \includegraphics[height=5.5cm]{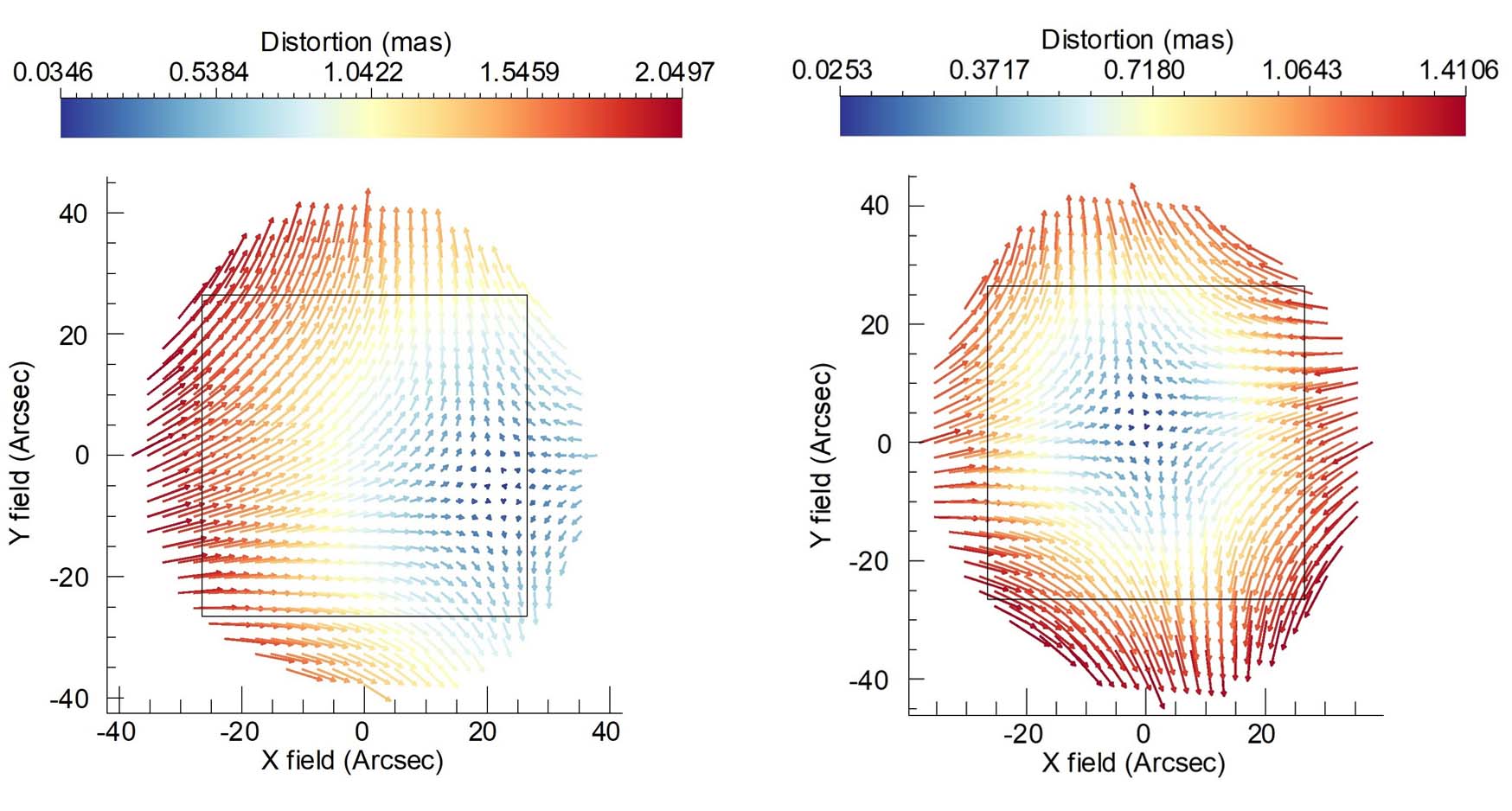}
    \\
  (a) \hspace{5.1cm} (b)
    \end{tabular}
\end{center}
\caption{(a): Worst case variation of the distortion pattern over the MICADO FoV with respect to the reference design. (b): same distortion map of (a) when the residual focal plane tilt is assumed to be compensated by the MCAO.}
\label{fig:21}       
\end{figure}

\begin{figure}[!h] 
\begin{center}
\begin{tabular}{c}
  \includegraphics[height=5.5cm]{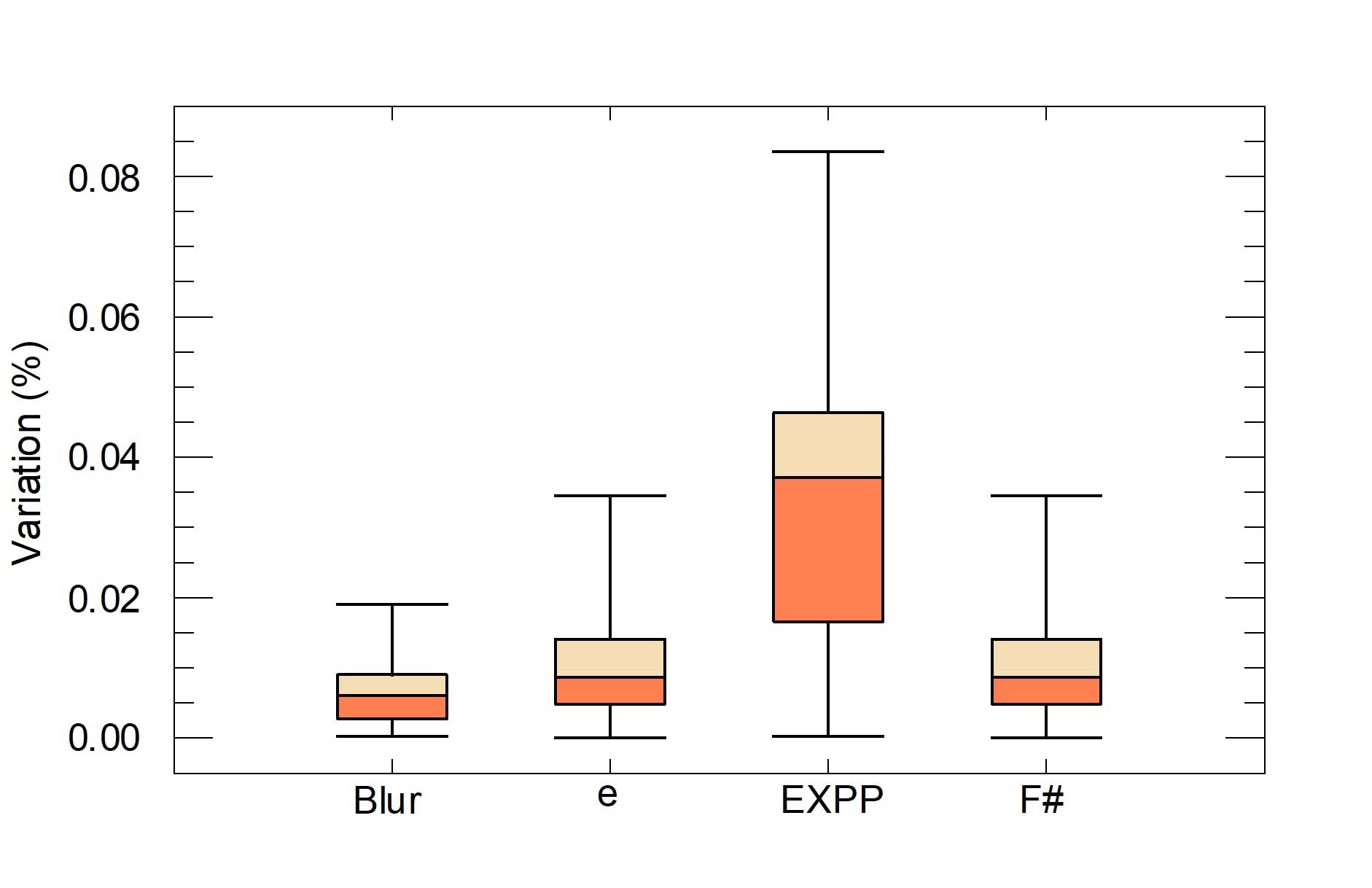}
    \end{tabular}
\end{center}
\caption{Full error simulation case: statistical distribution of the main optical parameters variation with respect to the reference design. Blur is the pupil blur variation; e is the pupil eccentricity variation; EXPP is the variation of the exit pupil position; F\# is the F-ratio variation. Box plots show minimum and maximum values and quartiles of aligned Monte Carlo trials parameters values.}
\label{fig:22}       
\end{figure}

\begin{figure}[!h]
\begin{center}
\begin{tabular}{c}
  \includegraphics[height=5.5cm]{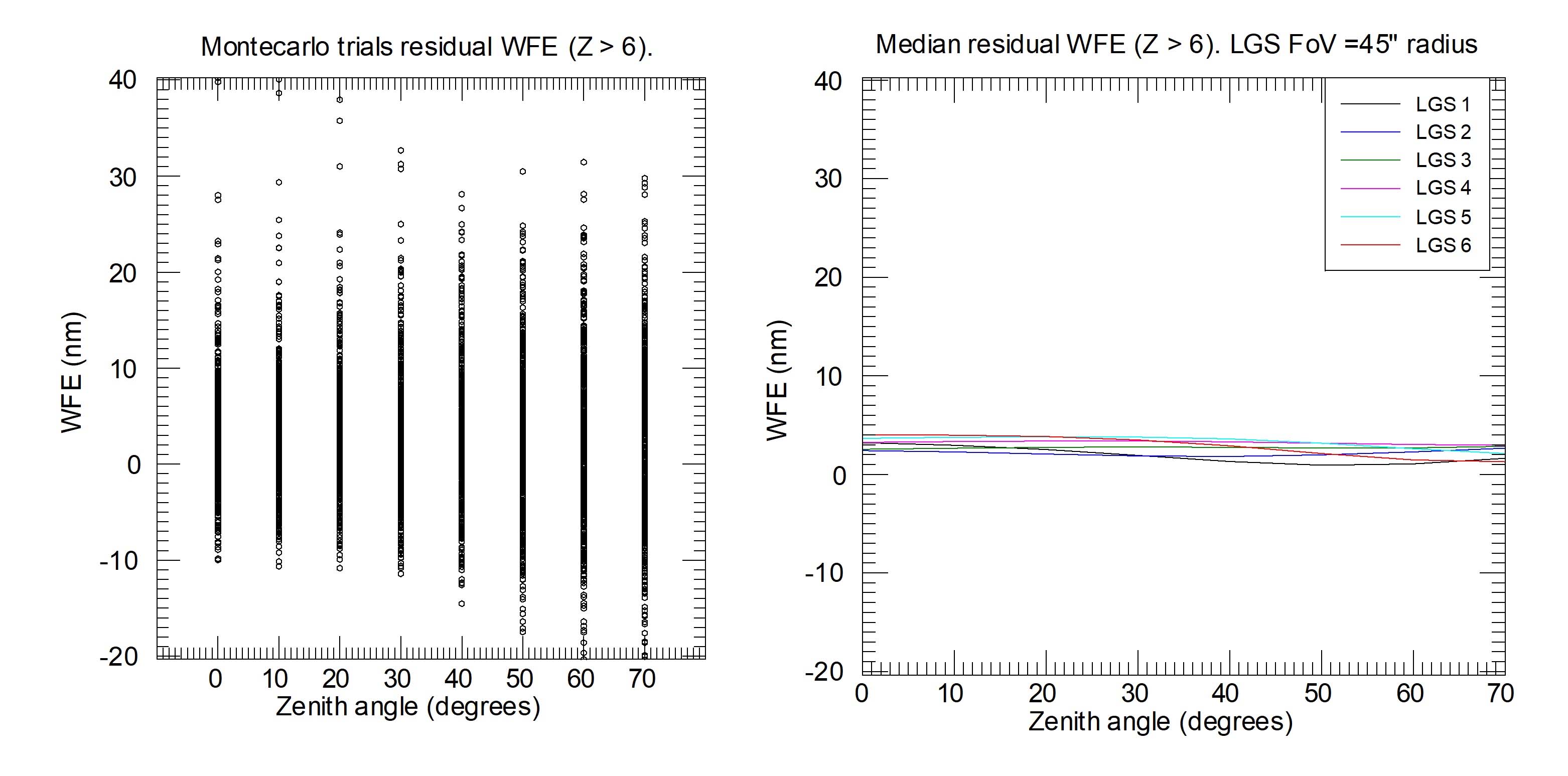}
      \\
  (a) \hspace{5.1cm} (b)
    \end{tabular}
\end{center}
\caption{(a): variation of RMS WFE (Tip-Tilt, defocus and astigmatism excluded) at the LGS Objective image plane with respect to the nominal value as a function of Zenith angle. (b): mean values of (a), different colours correspond to different positions around the 45 arcsec circle FoV of the LGSs.}
\label{fig:23}       
\end{figure}

The complete simulation of the MPO alignment by means of ZeRO includes all the errors described up to this point.  Figure~\ref{fig:18} shows the RMS WFE distribution of the Monte Carlo trials and its median map at the beginning of the procedure, when ZeRO has not achieved the alignment condition yet. It is the starting point, in terms of RMS WFE, considering the `all errors` simulation.
\\The major contribution to the RMS WFE, as discussed in the previous sub-section, is the defocus term. The errors which mostly contribute to the defocus are the uncertainties on surfaces curvature radii and chief ray angle of the artifical sources.\\Figure~\ref{fig:19} is the same set of Monte Carlo trials where the exit focal plane has been shifted to remove the mean defocus contribution to the RMS WFE. The median RMS WFE map within the scientific FoV is above the diffraction limit in J band ($\sim$89 $nm$) supporting the need of the ZeRO algorithm.
\pagebreak
\\Figures from~\ref{fig:20} to~\ref{fig:22} summarise the results of the ZeRO alignment. Table~\ref{tab:10} lists the extreme values of movements, as result of the algorithm output, that DoFs compensators had to apply through the Monte Carlo trials.\\
\begin{table}[!h]
\caption{Ranges of DoFs motion to compensate the misalignments given by the Monte Carlo trials in the full errors simulation.}
\label{tab:10}       
\begin{center}  
\begin{tabular}{lll}
\hline\noalign{\smallskip}
\textbf{Tilts} (M7-M8-M9-Dichroic)& \textbf{Axial shift} (M8)& \textbf{Tilts} (M11) \\
\noalign{\smallskip}\hline\noalign{\smallskip}
$\pm$5 $mrad$& $\pm$2.5 $mm$& $\pm$0.5 $mrad$\\
\end{tabular}
\end{center}  
\end{table}
\\Figure~\ref{fig:20} shows we are able to achieve a residual median RMS WFE of $\sim$20 $nm$ with a standard deviation of $\sim$15 $nm$. This is within the 40 $nm$ RMS WFE allocated in the error budget for manufacturing, assembly and integration of the instrument.
\\The RMS WFE distribution of Figure~\ref{fig:20} is similar to Figure~\ref{fig:14}, showing that the uncertainty about the optical surface irregularities are the major contributors to the RMS WFE variation with respect to the reference design. 
\\A major contribution to the variation of the geometric distortion, as shown in Figure~\ref{fig:21}(a), is given by the residual misalignment of the MAORY optical axis with respect to the de-rotator mechanical rotation axis. Since 55 $\mu m$ as centroid accuracy has been considered, the exit focal plane deviates from its nominal position within a range defined by the centroid accuracy.\\ As previously described, this focal plane drift will be permanently compensated by MCAO producing a real variation of the geometric distortion as shown on Figure~\ref{fig:21}(b). The MCAO should also remove the plate scale component of the geometric distortion further reducing the values of Figure~\ref{fig:21}(b) but this analysis is beyond the scope of this paper where the obtained results are within the specifications. 
\\We achieved a maximum value of 1.4 $mas$ of PSF blur which is below the allocated 2.4 $mas$ (Section~\ref{sec:3}).\\ Figure~\ref{fig:22} shows the difference between the reference design and the aligned
Monte Carlo trials considering different optical parameters. The variation of the main optical parameters is negligible meaning that the usage of the Zernike coefficients (RMS WFE), exit pupil and focal plane positions as targets for the ZeRO algorithm, is sufficient to keep the main optical parameters close to the nominal values.
\\
The ZeRO acts on some DoFs which are in common path with the LGS objective channel. A successful alignment of the MPO could be sub-optimal for the LGS objective. To evaluate the effect on the LGS image plane, the RMS WFE of 6 sodium LGSs has been extracted from the aligned Monte Carlo trials of the full errors ZeRO simulation. 
Fixing the LGS constellation at 45 arcsec angular radius which is the baseline asterism configuration, Figure~\ref{fig:23} shows the variation of the RMS WFE vs. Zenith angle with respect to the nominal value. Tip-Tilt, defocus and astigmatism (Z $<$ 6) have been excluded since they will be measured by the NGS WFSs within the MCAO loop. 
The LGS objective alignment procedure is beyond the scope of this paper. The goal of Figure~\ref{fig:23} is to demonstrates that if the LGS objective has been already aligned, after the ZeRO on the MPO, the optical quality variation of the LGS channel is negligible.
\section{Conclusions}
This paper describes an alignment method for complex optical systems. It is divided into two steps: mechanical alignment by LT and optical alignment by ZeRO. The results have been presented for the MAORY case, but they are generally valid for every optical system since the procedure is independent from the optical design.\\ 
Misalignments at the level of LT accuracy have been considered as well as different error sources that could increase the ZeRO uncertainty.\\
The ray-tracing simulations that have been run to verify the alignment method, also play a crucial role in defining the requirements on the geometry and number of sources for the calibration unit. \\
From a performance point of view, the optics parameters and irregularities must be within the tolerances which have been defined considering a possible compensation during the alignment. The HO surface irregularities cannot be compensated and they are the major contributors to the RMS WFE degradation.\\ The ray-tracing simulations verified that the ZeRO method is able to recover the nominal optical performance with very low residuals on RMS WFE, geometric distortions and main optical parameters.\\
The method is fully automatic and is a powerful tool to reduce the amount of time allocated to some activities during the AIV phase. Moreover, it could be used to actively control instrument optics at telescope site as part of routine re-alignment during maintenance or calibration operations.\pagebreak


\bibliography{Pap}   
\bibliographystyle{spiejour}   

\end{spacing}
\end{document}